\documentclass{article}
\usepackage[english]{babel}
\usepackage{amsmath}
\usepackage{amssymb}
\usepackage{example}
\usepackage{multibbl}
\usepackage{theorem}
\usepackage{relsize}
\usepackage{graphicx}
\usepackage{subfigure}
\graphicspath{{figss/}}
\usepackage{color}
\usepackage{pstricks,pst-all,pstricks-add,pst-plot,pst-eucl}
\usepackage{enumerate}
\usepackage[retainorgcmds]{IEEEtrantools}
\usepackage[labelfont=bf,labelsep=period,justification=raggedright]{caption}
\usepackage{enumerate}
\usepackage{soul}

\definecolor{brown}{rgb}{0.64,0.16,0.16}
\definecolor{ForestGreen}{rgb}{0.13,0.54,0.13}
\definecolor{purple}{rgb}{0.62,0.12,0.94}
\definecolor{DodgerBlue}{rgb}{0.11,0.56,0.98}
\definecolor{RoyalBlue}{rgb}{0.25,0.41,0.88}
\definecolor{B}{rgb}{0,0,1}
\definecolor{G}{rgb}{0,0.502,0}
\definecolor{R}{rgb}{1,0,0}
\definecolor{regenerative}{rgb}{0.78,0.86,0.94}
\definecolor{restorative}{rgb}{0.68,0.72,0.78}

\usepackage{framed}
\usepackage{lipsum}

\usepackage{ulem}

\newcommand {\MI}{\mathcal I}

\makeatletter
\renewcommand{\fnum@figure}{\small\textbf{\figurename~\thefigure}}
\makeatother

\theorembodyfont{\rm}

\title{\textbf{{A Balance Equation Determines a Switch in Neuronal Excitability}}}
\author{Alessio Franci$^{1,\#}$, Guillaume Drion$^{2,3\#}$, Vincent Seutin$^{3}$ \& Rodolphe Sepulchre$^{1,2,\ast}$\\
\small{$^1$INRIA Lille-Nord Europe, Orchestron project, 40 avenue Halley F 59650 Villeneuve d'Ascq, France}\\
\small{$^2$Department of Electrical Engineering and Computer Science and GIGA Research, University of Li\`ege, Li\`ege, Belgium.}\\
\small{$^3$Neurophysiology Unit, GIGA Neurosciences, University of Li\`ege, Li\`ege, Belgium.}\\
\small{$\#$These authors contributed equally to this work.}\\
\small{$\ast$ E-mail: R.Sepulchre@ulg.ac.be}}
\date{}

\begin{document}
\setulcolor{} 
\setstcolor{red}   
\sethlcolor{yellow}  

\maketitle
\newbibliography{main}
\section*{Abstract}

We use the qualitative insight of a planar neuronal phase portrait to detect an excitability switch
in arbitrary conductance-based models from a simple mathematical condition. The condition expresses a balance
between ion channels that provide a negative feedback at resting potential ({restorative} channels) and those
that provide a positive feedback at resting potential ({regenerative} channels). Geometrically, the condition imposes a transcritical bifurcation that rules the switch of excitability through the variation of a single 
physiological parameter. {Our analysis of} six different published conductance based models
always finds the transcritical bifurcation and the associated switch in excitability, which suggests that the mathematical predictions have a physiological relevance and that a same regulatory mechanism is potentially involved in the excitability and signaling of many neurons.

\section*{Author summary}

Understanding the changing electrophysiological signatures of neurons in
 different physiological and pharmacological conditions is a central
focus of experimental electrophysiology because a key component of cell
signaling in the nervous system. Computational modeling may assist
experimentalists in this quest by identifying core mechanisms and suggesting
pharmacological targets from a mathematical analysis of the model.  But
a successful interplay between experiments and mathematical predictions
requires new analysis tools adapted to the complexity of high-dimensional computational models nowadays available.
We use bifurcation theory to propose a mathematical condition that can detect an important
switch of neuronal excitability in arbitrary conductance-based neuronal models and we illustrate
its physiological relevance in six published state-of-the art models of different neurons.


\section*{Introduction}

Detailed computational conductance-based models have long demonstrated their ability to faithfully 
reproduce the variety of electrophysiological signatures that can be recorded from a {single} neuron
in varying physiological or pharmacological conditions.  But the predictive value of a computational model is
limited unless its analysis  sheds light on the core mechanisms at play behind a computer simulation.
Because conductance-based models are nonlinear dynamical models, their analysis often
requires a drastic reduction of dimension. The reduced model is amenable to the geometric methods
of dynamical systems theory, but the mathematical insight is often gained at the expense of physiological interpretability; hence the need for methodological
tools that can relate mathematical predictions of low-dimensional models to physiological predictions in detailed
conductance based models. 

In recent work \cite{main}{FRDRSE_SIAM}, we used phase plane analysis and dynamical bifurcation theory to characterize in {\it reduced-order}  neurodynamics models a switch of excitability that is consistent with many physiological observations. More precisely, a transcritical bifurcation governed by a single parameter {was shown to organize a switch from restorative excitability, extensively studied in most models inspired from the Hodgkin-Huxley model, to regenerative excitability whose distinct electrophysiological signature include spike latency, plateau oscillations, and afterdepolarizeation potentials.}.

The main contribution of the present paper is to show that this transcritical bifurcation, and the associated excitability switch, exist in a number of {\it high-dimensional} conductance-based models and that the resulting mathematical predictions have physiological relevance. Although purely mathematical in nature, the detection of the transcritical bifurcation relies on an ansatz that leads to a simple physiological interpretation: the switch of excitability is determined by a balance between {restorative} (those providing a negative feedback) and {regenerative} (those providing a positive feedback) ion channels at the resting potential. Because this {simple balance equation} can take many different physiological forms, it is potentially shared by very different neurons.

{We use the balance equation to provide an algorithm to trace the transcritical bifurcation in arbitrary conductance-based models.} We apply the algorithm to detailed conductance-based models of six neurons known to exhibit drastic changes in their electrophysiological signatures depending on environmental conditions: the squid giant axon \cite{main}{HODHUX}, the dopaminergic neuron \cite{main}{DRMASESE11}, the thalamic relay neuron \cite{main}{destexhe1998dendritic}, the thalamic reticular neuron \cite{main}{destexhe1996vivo}, the aplysia R15 model \cite{main}{rinzel1987dissection}, and the cerebellar granular cell \cite{main}{d2001theta}. In each case, the algorithm identifies a transcritical bifurcation that occurs {close to the nominal model} parameters and its predictions {are consistent with experimental} observations.

After defining a novel classification of ion channels based on their {restorative} or {regenerative} nature, we briefly review the planar model presented in \cite{main}{FRDRSE_SIAM} and how its transcritical bifurcation qualitatively captures the switch between {restorative} and {regenerative} excitability. As a generalization of this low-dimensional case, we mathematically construct the same bifurcation in generic conductance based models and derive the balance condition determining the {regenerative} or {restorative} nature of the model. This construction and its electrophysiological predictions are firstly illustrated on the squid giant axon. An algorithm for generic conductance-based models is subsequently derived and different models analysed.

\section*{Results}

\subsection*{{{Slow} restorative} and {{slow} regenerative} ion channels}

Conductance-based models of neurons describe the dynamic interaction between the membrane potential $V$ and - possibly many - gating variables that control the ionic flow through the membrane. The gating of ion channels occurs on many different timescales. However, gating timescales can be grouped in three families, according to their influence on neuronal excitablity \cite{main}{HILLE84}: 

\begin{enumerate}[(i)]
\item {\bf {Fast gating variables}}: These variables have a time-constant in the millisecond range. They generate the rapid regenerative upstroke of an action potential. Prominent representatives of this family are the activation gating variables of fast voltage-gated sodium channels (Na$_{\rm V}$1.1 to Na$_{\rm V}$1.9).

\item {\bf {Slow gating variables}}: These variables have a time constant 5 to 10 times larger than fast gating variables. They influence the spike initiation, downstroke, and the afterspike period. They are key players of neuronal excitability. Prominent representatives of this family are the activation gating variables of delayed rectifier potassium channels (K$_{\rm V}$1.1 to K$_{\rm V}$1.3, K$_{\rm V}$1.5 to K$_{\rm V}$1.8, K$_{\rm V}$2.1, K$_{\rm V}$2.2, K$_{\rm V}$3.1, K$_{\rm V}$3.2, K$_{\rm V}$7.1 to K$_{\rm V}$7.5, K$_{\rm V}$10.1) and the activation gating variables of all calcium channels (Ca$_{\rm V}$1.x, Ca$_{\rm V}$2.x, Ca$_{\rm V}$3.x).

\item {\bf {Ultra-Slow (adaptation)  variables}}: These variables gate too slowly to be strongly activated by single action potentials. They modulate neuronal excitability only over periods of many action potentials. Prominent representative of this family are the inactivation gating variables of transient calcium channels (Ca$_{\rm V}$2.x, Ca$_{\rm V}$3.x). Ultra-slow variables might also include non gating variables. For instance, the intracellular calcium concentration $[Ca^{2+}]_{in}$, which modulates the conductance of calcium-regulated channels.
\end{enumerate}

In view of their importance for neuronal excitability, we focus only on {slow} gating variables to classify ion channels: when the {slow} channel provides negative feedback to the membrane potential variation, we term the associated channel a \underline{{slow} restorative} ion channel. When the {slow} variable instead enhances a voltage variation by positive feedback, the associated ion channel is termed \underline{{slow} regenerative} (a characterization in terms of partial derivatives is postponed to the next sections). Ion channels that do not possess a {slow} gating variable are neither {restorative} nor {regenerative} and are called \underline{neutral}. Neutral ion channels solely regulate the ``quantity'' of excitability without affecting its ``quality''. 

Table \ref{TAB:IC} shows a classification of many known ion channels according to this criterion. Not surprisingly, potassium channels are the main representatives of {{slow }restorative} ion channels. By increasing the total outward current, their activation induces a negative feedback on membrane voltage variations that is responsible for neuron repolarization. On the other hand, physiologically described calcium channels are all {{slow} regenerative}. Their activation induces an increase of the total post-spike inward current, in contrast to potassium channels. This is the source, for instance, of afterdepolarization potentials (ADP). Interestingly, sodium channels can be either {restorative}, {regenerative}, or neutral according to their fast transient, resurgent, or persistent behavior, respectively. 

It is important to observe that the {restorative} (resp. {regenerative}) nature of channels is not solely linked to the outward (resp. inward) nature of the current. For instance, transient sodium channels {(although responsible for the regenerative spike upstroke)} are {{slow} restorative}, because their {slow} variable inactivates an inward current, inducing a negative-feedback on membrane potential variations. Similarly, potassium channels can be {{slow} regenerative} when their slow inactivation massively decreases the outward current, like in the case of A-type potassium channels. 

Although elementary, the classification above seems novel. It is motivated by the central message of this paper, that the balance between {regenerative} and {restorative} ion channels in  {slow timescale} determines its neuronal excitability type.{ In the remainder of the paper, we simply write restorative (resp. regenerative) for slow restorative (resp. slow regenerative) channels.}

\subsection*{{Restorative} and {regenerative} excitability in planar models}

Planar models - that only consist of two state variables - have been instrumental to study excitability since the early days of neurodynamics \cite{main}{fitzhugh61,rinzel1985excitation}. Empirical planar reductions of conductance based models only retain the (fast) voltage variable $V$ and one {slow} gating variable $n$. Fast gating variables are set to steady-state (i.e. their fast variation is approximated as instantaneous), adaptation variables are treated as slowly varying parameters, and the sole {gating} variable $n$ aggregates all {slow} variables, expressing each of them as a (curve-fitted) static function of $n$.

\begin{figure}[h!]
\centering
\includegraphics[width=0.8\textwidth]{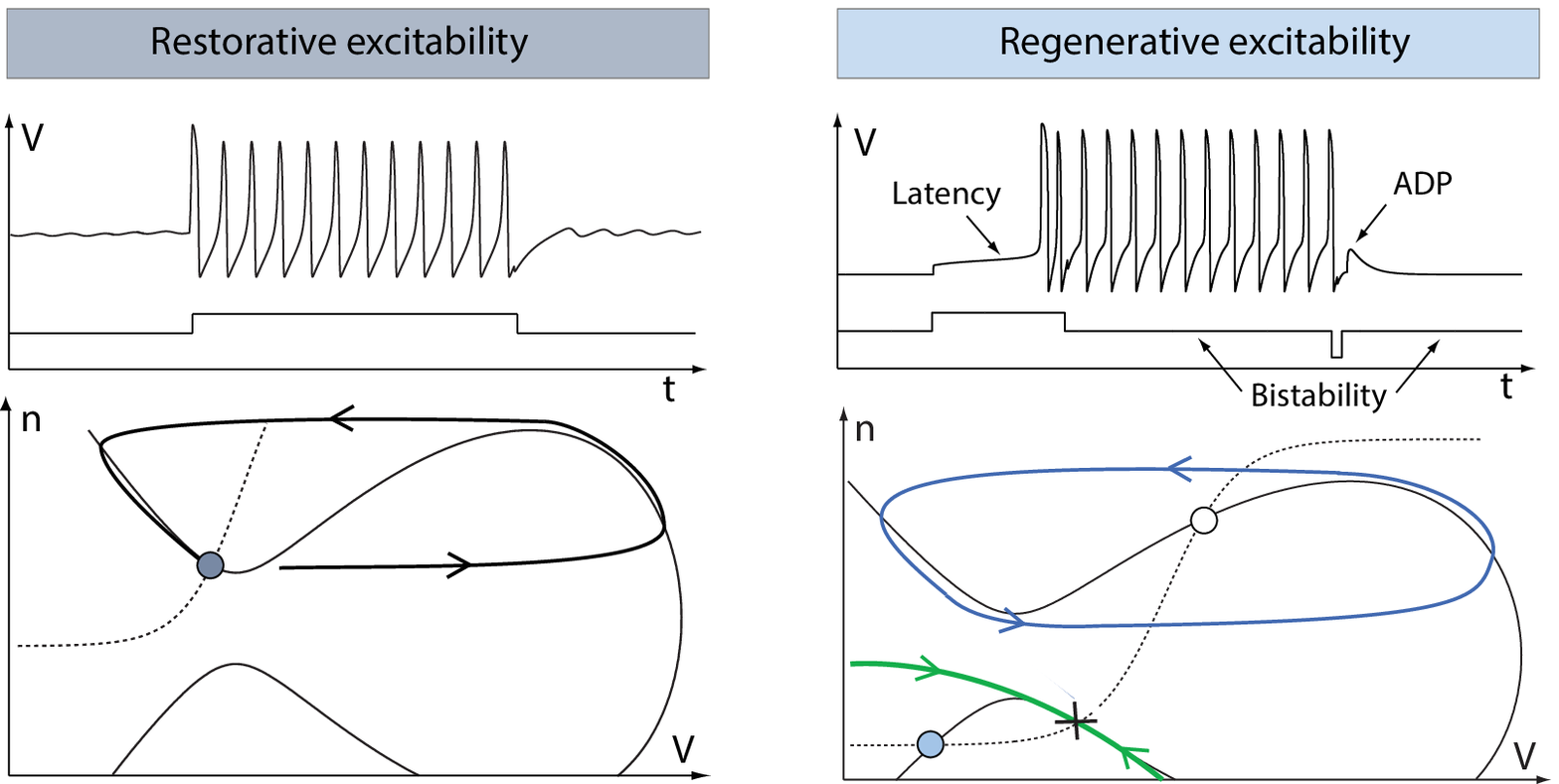}
\caption{The bottom figures illustrate the typical phase portrait of {restorative} (left) and {regenerative} (right) excitability. The dark (resp. light) blue circle denotes a stable {restorative} (resp. {regenerative}) steady state $(\bar V,\bar n)$. The thin full (resp. dashed) line is the voltage (resp. {slow} variable) nullcline. The saddle point in the right phase portrait is represented as the cross and its separatrix as the green oriented line. The stable limit cycle surrounding the unstable fixed point (represented as a circle) is represented by the blue oriented line. {The thick line in the left phase portrait represents the typical trajectory associated to the generation of an action potential.} The top figures illustrate the typical accompanying electrophysiological responses to step variations of current.}\label{FIG: SIAM exc types}
\end{figure}

Motivated by the phase portrait of such an empirical reduction of the Hodgkin-Huxley model augmented with a calcium channel \cite{main}{DRFRSESE_PLoS_sub}, our recent study \cite{main}{FRDRSE_SIAM} explores the neuronal excitability of the planar model
\begin{IEEEeqnarray}{rCl}\label{EQ: SIAM planar model}
\dot V&=&V-\frac{V^3}{3}-n^2+I_{app}\IEEEyessubnumber\\
\dot n&=&\varepsilon(n_\infty(V-V_0)+n_0-n)\IEEEyessubnumber
\end{IEEEeqnarray}
whose phase portraits are reproduced in Fig. \ref{FIG: SIAM exc types} for two distinct values of the parameter $n_0$ (an indirect {representation} of the calcium conductance in the high-dimensional model). The parameter $\varepsilon>0$ {characterizes} the time-scale separation between $V$ and $n$. The function $n_\infty(\cdot)$ has the standard sigmoid shape of conductance-based models and $V_0$ is the half-activation potential. {The typical step responses of} (\ref{EQ: SIAM planar model}) {are also reproduced in Fig.} \ref{FIG: SIAM exc types}.

{The spike generation mechanism in the phase portrait of Fig.} \ref{FIG: SIAM exc types} left is {reminiscent of the} of FitzHugh-Nagumo model {and of the physiologically grounded planar reduction of Hodgkin-Huxley model by Rinzel} \cite{main}{rinzel1985excitation}. It is associated to a reversible and sudden switch from rest to firing and has extensively been studied, with finer distinctions depending on the mathematical nature of the underlying bifurcation \cite{main}{ermentrout1996type}. {For further reading, see} \cite{main}{Rinzel:1989:ANE:94605.94613},\cite[Section 7.1.3]{main}{IZHIKEVICH2007} and references therein.


The phase portrait in Fig. \ref{FIG: SIAM exc types} right is in sharp contrast in that the electrophysiological response to a current input exhibits spike latency, plateau oscillations, and after depolarization potential (ADP). This specific signature, experimentally observed in many families of neurons, is fundamentally associated to the bistability illustrated in the phase portrait: namely, the robust coexistence of two stable attractors (a hyperpolarized resting potential and a limit cycle of periodic action potentials) and a saddle-separatrix that sharply separates their basins of attraction.
The time evolution shown in the top figure is a consequence of this phase portrait and cannot be observed in FitzHugh-Nagumo  like phase portraits.
{The distinction between the two phase portraits{, the associated excitability types, and their relation with Hodgkin's excitability classification} \cite{main}{hodgkin1948local} are further discussed in \cite{main}{FRDRSE_SIAM} and later in the paper.}

\begin{figure}
\centering
\includegraphics[width=0.7\textwidth]{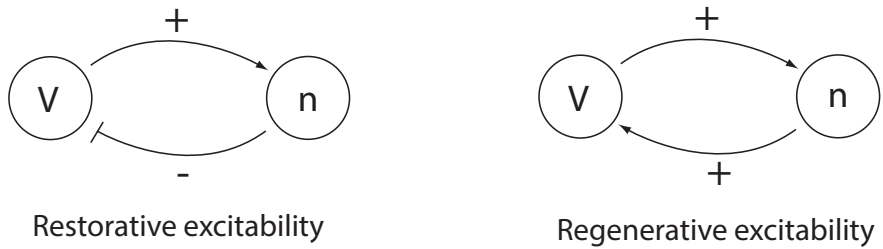}
\caption{Block diagrams representation of {restorative} and {regenerative} excitability in planar models.}\label{FIG: comp/coop block diagram}
\end{figure}

A simple mathematical distinction between the two phase portraits shown in Fig. \ref{FIG: SIAM exc types} is drawn from the Jacobian linearization of the model at the stable resting point $(\bar V,\bar n)$:
\begin{equation*}
J=\left.\left(\begin{array}{cc}
\displaystyle{\frac{\partial\dot V}{\partial V}} & \displaystyle{\frac{\partial\dot V}{\partial n}}\\
&\\
\displaystyle{\frac{\partial\dot n}{\partial V}} & \displaystyle{\frac{\partial\dot n}{\partial n}}
\end{array} \right)\right|_{(\bar V,\bar n)}=
\left(\begin{array}{cc}
1-\bar V^2 & -2\bar n\\
\varepsilon \underbrace{\frac{\partial n_\infty}{\partial V}(\bar V-V_0)}_{>0} & -\varepsilon
\end{array}\right)\,.
\end{equation*}
The product of the partial derivatives $\frac{\partial\dot V}{\partial n}\frac{\partial\dot n}{\partial V}=-2\bar n\,\varepsilon \frac{\partial n_\infty}{\partial V}(\bar V-V_0)$ is negative on the left phase portrait {($\bar n>0$)}{, capturing the restorative nature of the gating variable,}  whereas it is positive on the right phase portrait {($\bar n<0$)}{, translating the regenerative nature of the gating variable}. This difference is schematized in the block diagrams of Fig. \ref{FIG: comp/coop block diagram}. {Accordingly, excitability in planar models is called restorative (resp. regenerative) when the gating variable provides negative (resp. positive) feedback close to the resting point:}
\begin{center}
\begin{tabular}{c|c}
\colorbox{restorative}{\phantom{aaaaaaaaaa} Planar estorative excitability \phantom{aaaaaaaaaa}} & \colorbox{regenerative}{\phantom{aaaaaaaaaa} Planar regenerative excitability \phantom{aaaaaaaaaa}}\\
{\color{white}.}&{\color{white}.}\\
The model is said to be {restorative} at steady state if: & The model is said to be {regenerative} at steady state if:\\
{\color{white}.}&{\color{white}.}\\
$\displaystyle{\left.\frac{\partial\dot V}{\partial n} \frac{\partial n_\infty}{\partial V}\right|_{SS}}<0$
&
$\displaystyle{\left.\frac{\partial\dot V}{\partial n} \frac{\partial n_\infty}{\partial V}\right|_{SS}}>0$
\end{tabular}
\end{center}

\begin{figure}[h!]
\center
\includegraphics[width=0.7\textwidth]{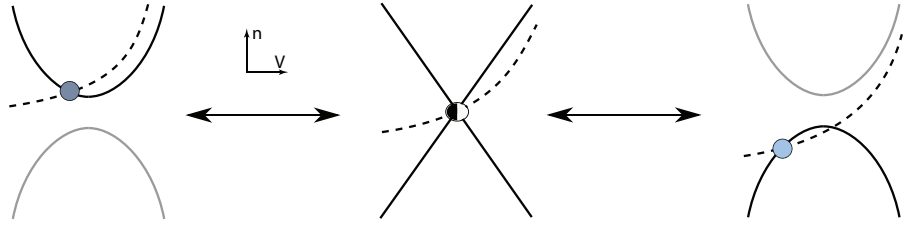}
\caption{A continuous deformation from the {restorative} phase portrait of Fig. \ref{FIG: SIAM exc types} left to the {regenerative} phase portrait of Fig. \ref{FIG: SIAM exc types} right involving a transcritical bifurcation \cite[Section 3.2]{main}{STROGATZchaos} determined by the algebraic conditions (\ref{EQ: planar bal eq}) and (\ref{EQ: bif condition}). The dark blue circle represents a {restorative} stable steady-state, the light blue circle a {regenerative} stable steady-state, and the half-filled circle represents the transcritical bifurcation which separates the {restorative} and {regenerative} regimes.}\label{EQ: geom trans constr}
\end{figure}

The planar model (\ref{EQ: SIAM planar model}) can smoothly switch from {restorative} to {regenerative} excitability, with a transition occurring for $\bar n=0$, or, in algebraic terms,
\begin{IEEEeqnarray}{rCl}\label{EQ: planar bal eq}
\frac{\partial\dot V}{\partial n} \frac{\partial n_\infty}{\partial V} &=& 0
\end{IEEEeqnarray}
A convenient way to algorithmically track this excitability switch is to use bifurcation analysis and to impose that the critical condition (\ref{EQ: planar bal eq}) coincides with a bifurcation of the model, which imposes the additional algebraic condition
\begin{IEEEeqnarray}{rCl}\label{EQ: bif condition}
\det J &=& \colorbox{white}{$\displaystyle{-\varepsilon\left(\frac{\partial \dot V}{\partial V} +\frac{\partial\dot V}{\partial n} \frac{\partial n_\infty}{\partial V}\right)} $} \nonumber\\
&=& 0.
\end{IEEEeqnarray}
Simultaneously imposing (\ref{EQ: planar bal eq}) and (\ref{EQ: bif condition}) implies
\begin{IEEEeqnarray}{rCl}\label{EQ: planar ansatz}
\frac{\partial \dot V}{\partial V}&=&0,
\end{IEEEeqnarray}
which, in geometrical terms, corresponds to the transcritical bifurcation obtained for $I_{app}=\frac{2}{3}$ and illustrated in Fig. \ref{EQ: geom trans constr}.

\begin{figure}[h!]
\centering
\includegraphics[width=0.7\textwidth]{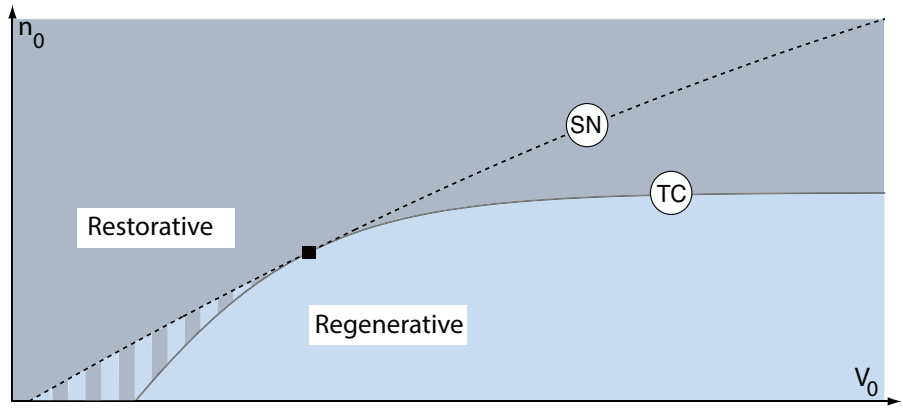}
\caption{ Excitability types in model (\ref{EQ: SIAM planar model}). SN denotes the saddle-node bifurcation, TC the transcritical bifurcation. $\blacksquare$: Pitchfork bifurcation organizing center. Varying $n_0$ and $V_0$  the model switches between excitability types.}\label{FIG: SIAM PC}
\end{figure}

The theory of bifurcation unfolding is further exploited in \cite{main}{FRDRSE_SIAM} in order to classify all excitability types associated to the planar model in Fig. \ref{FIG: SIAM exc types}. This analysis results in five different types of excitability obtained by varying the two parameters $(V_0,n_0)$ around the singular phase portrait of Fig. \ref{EQ: geom trans constr}, center. The parameter $n_0$ acts in particular as the sole regulator of the balance between {regenerative} and {restorative} excitability by shifting the $n$-nullcline up and down: a positive $n_0$ corresponds to a phase portrait as in Fig. \ref{EQ: geom trans constr} left, whereas the phase portrait of Fig. \ref{EQ: geom trans constr} right is obtained for sufficiently negative $n_0$. {An early graphical manifestation of such phase portraits in conductance-based models is in} \cite[Figs. 17, 18, and 19]{main}{RUTE04}. {Figure} \ref{FIG: SIAM PC} {delineates the two types of excitability in a two parameter chart}. It contains the {two} types of excitability discussed above. {The transition from restorative to regenerative excitability is always through a transcritical bifurcation. In addition, some paths traverse a small {mixed} region where a down regenerative steady state and an up restorative steady state coexist.}

Our main contribution in the present paper is to show that the diagram in Fig. \ref{FIG: SIAM PC} is not an artifact of planar reduction but captures excitability transitions that can algorithmically be tracked in conductance-based models of arbitrary dimension {by imposing a simple physiologically relevant algebraic condition}.

\subsection*{{Restorative} and {regenerative} excitability in conductance based models}

We start by grouping gating variables of a given conductance-based model according to their time scales. The family $\mathcal G_F=\{m_{Na,f},\ m_{Na,p},\ m_{K,A}, \ldots\}$ collects fast gating variables. The gating variable $x^f\in[0,1]$ denotes a generic member of this family. Similarly, the family $\mathcal G_{S}=\{h_{Na,f},\ m_{K,DR},\ m_{Ca,L},\ldots\}$ collects {slow} gating variables $x^s$, whereas $\mathcal G_{A}=\{h_{Ca,T},\ h_{Na,R},\ m_{K,M},\ldots\}$ collects adaptation variables $x^a$. For a given ion channel type $i$, the standard notation $m_i$ (resp. $h_i$) is adopted for the activation (resp. inactivation) gating variable of the associated ionic current $I_i$. With these notations, a general neuron conductance-based model reads
\begin{IEEEeqnarray}{rCl}\label{EQ: generic cb model}
C_m\dot V&=&-\sum_{i}\bar g_i m_i^{a_i} h_i^{b_i} (V-E_i)+I_{app},\IEEEyessubnumber\\
\tau_{x^f}(V)\dot x^f&=&(x^f_\infty(V)-x^f),\IEEEyessubnumber\\
\tau_{x^s}(V)\dot x^s&=&(x^s_\infty(V)-x^s),\IEEEyessubnumber\\
\tau_{x^a}(V)\dot x^a&=&(x^a_\infty(V)-x^a),\IEEEyessubnumber
\end{IEEEeqnarray}
where the sum in (\ref{EQ: generic cb model}a) is over all ion channels in the model, and (\ref{EQ: generic cb model}b),(\ref{EQ: generic cb model}c),(\ref{EQ: generic cb model}d) hold for all the associated fast, {slow}, and adaptation variables, respectively. The activation (resp. inactivation) functions $x^f_\infty,\,x^s_\infty$ are strictly monotone increasing (resp. decreasing) sigmoids. In the forthcoming analysis, all adaptation variables are treated as constant parameter, that is their slow evolution is neglected.

We will detect a switch from {restorative} to {regenerative} excitability by mimicking the two-dimensional algorithm of the previous section. We first impose the bifurcation condition $\det J=0$, where $J$ denotes the Jacobian of the subsystem (\ref{EQ: generic cb model}a),(\ref{EQ: generic cb model}b),(\ref{EQ: generic cb model}c). The algebraic condition writes
\begin{IEEEeqnarray}{rCl}\label{EQ: generic det split}
\frac{\partial\dot V}{\partial V}+\mathlarger{\mathlarger{\sum}}_{x^f} \frac{\partial\dot V}{\partial x^f} \frac{\partial x^f_\infty}{\partial V}+\mathlarger{\mathlarger{\sum}}_{x^s} \frac{\partial\dot V}{\partial x^s} \frac{\partial x^s_\infty}{\partial V}=0,
\end{IEEEeqnarray}
{where the sums are over all fast and  {slow} variables, respectively}. The particular form of equation (\ref{EQ: generic det split}) is a direct consequence of the specific structure of conductance-based models, that is, parallel interconnection of two-dimensional feedback loops involving the voltage dynamics (\ref{EQ: generic cb model}a) and one of the gating variable dynamics (\ref{EQ: generic cb model}b),(\ref{EQ: generic cb model}c).\\
As for the planar model (\ref{EQ: SIAM planar model}), we track the switch between {restorative} and {regenerative} excitability by imposing the high-dimensional equivalent of the balance condition (\ref{EQ: planar bal eq}). We therefore look for solutions of (\ref{EQ: generic det split}) satisfying the ansatz
\begin{IEEEeqnarray}{rCl}\label{EQ: generic bal eq}
\mathlarger{\mathlarger{\sum}}_{x^s} \frac{\partial\dot V}{\partial x^s} \frac{\partial x^s_\infty}{\partial V} &=& 0.
\end{IEEEeqnarray}
The two conditions (\ref{EQ: generic det split}),(\ref{EQ: generic bal eq}) now imply
\begin{IEEEeqnarray}{rCl}\label{EQ: generic ansatz}
\det J_f &=& 0,
\end{IEEEeqnarray}
{where $J_f$ denotes the Jacobian of the fast subsystem} (\ref{EQ: generic cb model}a),(\ref{EQ: generic cb model}b).
We show in Supplementary Material S1 that the corresponding bifurcation is necessarily transcritical \cite[Section 3.2]{main}{STROGATZchaos}.

The singularity {condition} (\ref{EQ: generic ansatz}) is the high-dimensional counterpart of the $V$-nullcline self-intersection in the planar model. It reflects the geometric nature of the transcritical bifurcation, that is, a robust geometrical object that exists independently of the timescale separation and persists in the singular limit of an infinite timescale separation, regardless of the system dimension. Our ansatz makes the proposed analysis completely robust against the model time constants. The time constants are only used to classify the gating variables in the three physiological groups.

We split $\mathcal G_S$ in the two subfamilies $\mathcal G_{S+}$, which contains {\it {regenerative}} {slow} gating variables $x^{s+}$, and $\mathcal G_{S-}$, which contains {\it {restorative}} {slow} gating variables $x^{s-}$. The balance condition (\ref{EQ: generic bal eq}) is then rewritten as
\begin{IEEEeqnarray}{rCl}\label{EQ: fundamental}
{\underbrace{\mathlarger{\mathlarger{\sum}}_{x^{s+}}\ \overbrace{\left.\frac{\partial \dot V}{\partial x^{s+}}\frac{\partial x^{s+}_\infty}{\partial V}\right|_{TC}}^{\colorbox{regenerative}{$\mathlarger{>0}$}} }_{\colorbox{regenerative}{{regenerative} gates}}\quad\quad +\quad\quad \underbrace{\mathlarger{\mathlarger{\sum}}_{x^{s-}}\ \overbrace{\left.\frac{\partial \dot V}{\partial x^{s-}}\frac{\partial x^{s-}_\infty}{\partial V}\right|_{TC}}^{\colorbox{restorative}{$\mathlarger{<0}$}} }_{\colorbox{restorative}{{restorative} gates}}=0}
\end{IEEEeqnarray}
to express a balance between {restorative} and {regenerative} ion channels. It is the high-dimensional counterpart of (\ref{EQ: planar bal eq}) and it provides a rigorous high-dimensional generalization of {restorative} and {regenerative} excitability:

\begin{center}
\begin{tabular}{c|c}
\colorbox{restorative}{\phantom{aaaaaaaaaaaaa} {Restorative} excitability \phantom{aaaaaaaaaaaaa}} & \colorbox{regenerative}{\phantom{aaaaaaaaaaaaa} {Regenerative} excitability \phantom{aaaaaaaaaaaaa}}\\
{\color{white}.}&{\color{white}.}\\
The model is said to be {restorative} at steady state if: & The model is said to be {regenerative} at steady state if:\\
{\color{white}.}&{\color{white}.}\\
$\displaystyle{\left.\left(\mathlarger{\mathlarger{\sum}}_{x^{s+}}\frac{\partial \dot V}{\partial x^{s+}}\frac{\partial x^{s+}_\infty}{\partial V}+\mathlarger{\mathlarger{\sum}}_{x^{s-}}\frac{\partial \dot V}{\partial x^{s-}}\frac{\partial x^{s-}_\infty}{\partial V}\right)\right|_{SS}<0}$
&
$\displaystyle{\left.\left(\mathlarger{\mathlarger{\sum}}_{x^{s+}}\frac{\partial \dot V}{\partial x^{s+}}\frac{\partial x^{s+}_\infty}{\partial V}+\mathlarger{\mathlarger{\sum}}_{x^{s-}}\frac{\partial \dot V}{\partial x^{s-}}\frac{\partial x^{s-}_\infty}{\partial V}\right)\right|_{SS}>0}$\\
{\color{white}.}&{\color{white}.}\\
\includegraphics[width=0.25\textwidth]{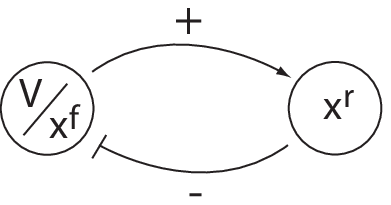} & \includegraphics[width=0.25\textwidth]{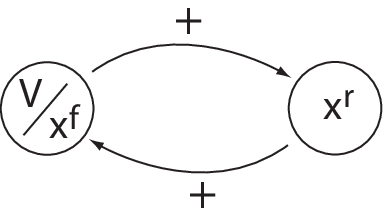}\\ 
{\color{white}.}&{\color{white}.}\\
\end{tabular}
\end{center}

The insight provided by the planar model of the previous section predicts that the switch of excitability detected by the balance equation (\ref{EQ: fundamental}) will lead to the accompanying distinct electrophysiological signatures of~Fig.~\ref{FIG: SIAM exc types}.

\subsection*{Tracking excitability switches in the squid giant axon}

The Hodgkin-Huxley (HH) model \cite{main}{HODHUX} provides a non-physiological, but historical and experimentally verified tutorial for tracking a switch of excitability in conductance based models. The model reads
\begin{IEEEeqnarray}{rCl}\label{EQ: HH model}
C\dot V &=&- \bar g_Kn^4(V-V_K) - \bar g_{Na}m^3h(V-V_{Na})- g_l(V-V_l) + I_{app},\IEEEyessubnumber\\
\tau_m(V)\dot m&=&(m_{\infty}(V)-m),\IEEEyessubnumber\\
\tau_h(V)\dot h&=&(h_{\infty}(V)-h),\IEEEyessubnumber\\
\tau_n(V)\dot n&=&(n_{\infty}(V)-n),\IEEEyessubnumber
\end{IEEEeqnarray}
where $m$ is the fast sodium channel activation while the sodium channel inactivation $h$ and the potassium channel activation $n$ are the {slow} gating variables. We set all time constants to one, because this simplification has no effects on the algebraic conditions (\ref{EQ: generic bal eq}) and (\ref{EQ: generic ansatz}). The Jacobian of (\ref{EQ: HH model}) reads
\begin{equation}\label{EQ: HH Jacobian}
J=\left[\begin{array}{cc|cc}
{\color{white}\displaystyle{\frac{a}{.}}}\frac{\partial\dot V}{\partial V} & \frac{\partial\dot V}{\partial m} & \frac{\partial\dot V}{\partial h} & \frac{\partial\dot V}{\partial n}{\color{white}\displaystyle{\frac{a}{.}}}\\
{\color{white}\displaystyle{\frac{a}{.}}}\frac{\partial m_{\infty}}{\partial V} & -1 & 0 & 0{\color{white}\displaystyle{\frac{a}{.}}}\\
\hline
{\color{white}\displaystyle{\frac{a}{.}}}\frac{\partial h_{\infty}}{\partial V} & 0 & -1 & 0{\color{white}\displaystyle{\frac{a}{.}}}\\
{\color{white}\displaystyle{\frac{a}{.}}}\frac{\partial n_{\infty}}{\partial V} & 0 & 0 & -1{\color{white}\displaystyle{\frac{a}{.}}}
\end{array}\right].
\end{equation}
The upper-left block is the Jacobian of the fast $(V,m)$ subsystem. Imposing the singularity condition (\ref{EQ: generic ansatz}) yields
\begin{IEEEeqnarray}{rCl}\label{EQ: HH ansatz}
\left.\frac{\partial\dot V}{\partial V}\right|_{TC}+\left.\frac{\partial\dot V}{\partial m}\frac{\partial m_{\infty}}{\partial V}\right|_{TC}&=&0,\IEEEyessubnumber
\end{IEEEeqnarray}
while the balance equation (\ref{EQ: fundamental}) reads
\begin{IEEEeqnarray}{rCl}\label{EQ: HH fundamentalbis}
\left.\frac{\partial\dot V}{\partial n}\frac{\partial n_{\infty}}{\partial V}\right|_{TC}+\left.\frac{\partial\dot V}{\partial h}\frac{\partial h_{\infty}}{\partial V}\right|_{TC}&=&0\IEEEyessubnumber
\end{IEEEeqnarray}
Note that (\ref{EQ: HH ansatz}) and (\ref{EQ: HH fundamentalbis}) imply the bifurcation condition $\det J=0$ in (\ref{EQ: HH Jacobian}).

At first sight, the balance condition (\ref{EQ: HH fundamentalbis}) cannot be satisfied because both sodium and potassium channels are {restorative} channels according to their corresponding kinetics in the model, and in agreement with our proposed classification. This is consistent with the fact that the excitability of the HH model is always {restorative} in physiological conditions.

However, it was long recognized \cite{main}{moore1959excitation} that potassium channels can generate an inward current at steady-state if the extracellular $K^+$ concentration is sufficiently large. Indeed, any change in extracellular potassium concentration induces a change in the potassium reversal potential, as expressed by the Nernst equation. This suggests to use the potassium reversal potential $V_K$ as a bifurcation parameter in HH model in order to satisfy the balance equation
\begin{IEEEeqnarray}{rCl}\label{EQ: HH fundamental}
{\underbrace{\overbrace{\left.\frac{\partial \dot V}{\partial n}\frac{\partial n_\infty}{\partial V}\right|_{TC}}^{\colorbox{regenerative}{$\mathlarger{!>0}$}} }_{\colorbox{regenerative}{$n$: {regenerative} gating}}\quad\quad +\quad\quad \underbrace{\overbrace{\left.\frac{\partial \dot V}{\partial h}\frac{\partial h_\infty}{\partial V}\right|_{TC}}^{\colorbox{restorative}{$\mathlarger{<0}$}} }_{\colorbox{restorative}{$h$: {restorative} gating}}=0}
\end{IEEEeqnarray}
where potassium now acts as a {regenerative} gating variable provided that $V_K>V_{SS}$. Physiologically, condition (\ref{EQ: HH fundamental}) imposes that the potassium Nernst potential is large enough for the regenerativity of the potassium activation to balance the {restorative} effects of the sodium current inactivation.

\begin{figure}
\centering
\includegraphics[width=0.87\textwidth]{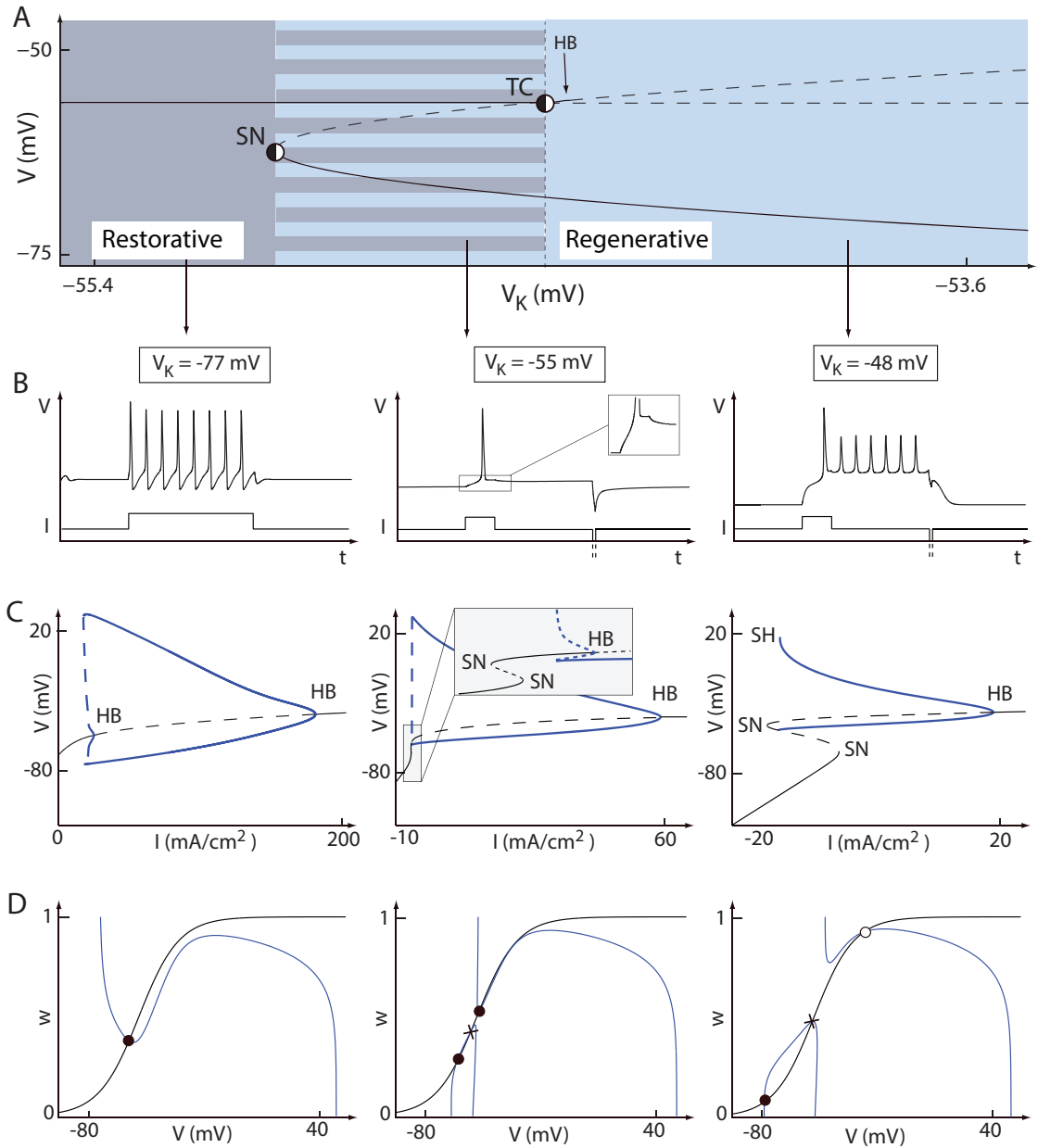}
\caption{{\bf Variations of the potassium reversal potential $V_K$ induce excitability switches in the Hodgkin-Huxley model}. {\bf A.} Bifurcation diagram of the HH model with $V_K$ as the bifurcation parameter. TC denotes a transcritical bifurcation, SN a saddle-node bifurcation, HB a Hopf bifurcation. {Branches of} stable fixed points are represented as solid line, {whereas branches of} saddle points and unstable points as dashed lines. {\bf B.} Electrophysiological responses of the model for three different values of $V_K$, corresponding to three different excitability types (restorative, {mixed}, and regenerative, from left to right). {\bf C.} Bifurcation diagrams with the applied current as the bifurcation parameter for the same three values of $V_K$ as in {\bf B}. Black (resp. blue) full lines represent {branches} of stable steady-states (resp. limit cycles), black dashed lines {branches} of saddle and unstable steady-states. {Branches of unstable limit cycle are drawn as dashed blue lines}. HB denotes a Hopf bifurcation, SN a saddle-node bifurcation, and SH a saddle-homoclinic bifurcation. {\bf D.} Phase portraits of reduced HH model proposed by Rinzel in \cite{main}{rinzel1985excitation} for the same three values of $V_K$ as in {\bf B,C}. Blue full lines denote the $V$-nullclines and black full lines the $w$-nullclines, where $w$ denotes the {slow} variable of the reduced model. Filled circles denote stable steady-states, crosses saddle points, and circles unstable steady-states.}\label{FIG: TC in HH}
\end{figure}

The two conditions (\ref{EQ: HH ansatz}) and (\ref{EQ: HH fundamental}) can numerically be solved to determine the critical values $V_{K}^c$ and $V^c$. The value of the applied current at the transcritical bifurcation is then determined from (\ref{EQ: HH model}a), which gives
\begin{equation*}
I_{app}^c= \bar g_Kn^4_\infty(V^c)(V^c-V_K^c) + \bar g_{Na}m^3_\infty(V^c)h_\infty(V^c)(V^c-V_{Na})+ g_l(V^c-V_l).
\end{equation*}
The numerical bifurcation diagram in Fig. \ref{FIG: TC in HH}A confirms the transcritical bifurcation at $V_K^c$. That bifurcation diagram is drawn by varying $V_K$ together with applied current $I_{app}$, following the affine reparametrization described in Supplementary Material S1. More precisely,
$$I_{app}(V_K)=I_{app}^c-\bar g_K n_\infty^4(V^c)(V_K-V_{K}^c).$$
Mathematically, this reparametrization imposes one of the defining conditions of the transcritical bifurcation. Physiologically, its effect is to keep the net current constant at steady-state $V^c$ ({\it i.e.} $I_{ion}(V_K)+I_{app}(V_K)=0$): as $V_K$ is varied, the observed switch in the excitability type does not rely on changes in the net current across the membrane, but solely on changes in its dynamical properties.

The bifurcation diagram in Fig. \ref{FIG: TC in HH}A provides informations on the model excitability also far from the transcritical values. For highly hyperpolarized $V_K$, the model is purely {restorative} and exhibits the typical excitable behavior of the original Hodgkin-Huxley model \cite[Figure 8]{main}{FRDRSE_SIAM}.
As $V_K$ is increased, a stable {regenerative} steady-state is born in a saddle-node bifurcation. At this transition, the system switches to a {mixed} excitability type. Short current pulses let the system switch between the depolarized {restorative} steady state and the hyperpolarized {regenerative} stable steady state (Fig. \ref{FIG: TC in HH}B,middle). {The associated bifurcation diagram and phase portrait are reproduced in Fig.} \ref{FIG: TC in HH}C,D,middle. Finally, further increase of $V_K$ let the {restorative} steady state exchanges its stability with a ({regenerative}) saddle at the transcritical bifurcation {(and, soon after, lose stability in a Hopf bifurcation)} and the system switches to {regenerative} excitability. The {regenerative} steady state coexists in this case with the spiking limit cycle attractor. Current pulses switch the systems between the two attractors (Fig. \ref{FIG: TC in HH}B,right). {The associated bifurcation diagram and phase portrait are reproduced in Fig.} \ref{FIG: TC in HH}C,D,right.

The same qualitative excitability switch was described by Rinzel in \cite{main}{rinzel1985excitation}, who linked the appearance of a bistable behavior to the inward nature of potassium current at steady-state for sufficiently depolarized $V_K$. {\it In vitro} recordings of the squid giant axon with isotonic extracellular $K^+$ concentration show the same transition \cite{main}{moore1959excitation}.

\subsection*{Tracking excitability switches in conductance-based models}

The mathematical analysis of the previous sections follows an algorithm that allows to detect a transcritical bifurcation in generic conductance based models of arbitrary dimension and to track associated excitability switches. The steps of the algorithm are summarized in Table \ref{TAB: TC algo}. {For simplicity and conciseness, we restrict our attention to the modulation of only one regenerative ionic current at a time. However, a similar algorithm can be written for an arbitrary modulation of ionic currents {(by variation of maximal conductance(s), adaptation variable(s), or reverse potential(s))} that brings the model to the balance
expressed in} (\ref{EQ: fundamental}).

\begin{figure}[h!]
\begin{center}
\includegraphics[width=0.7\linewidth]{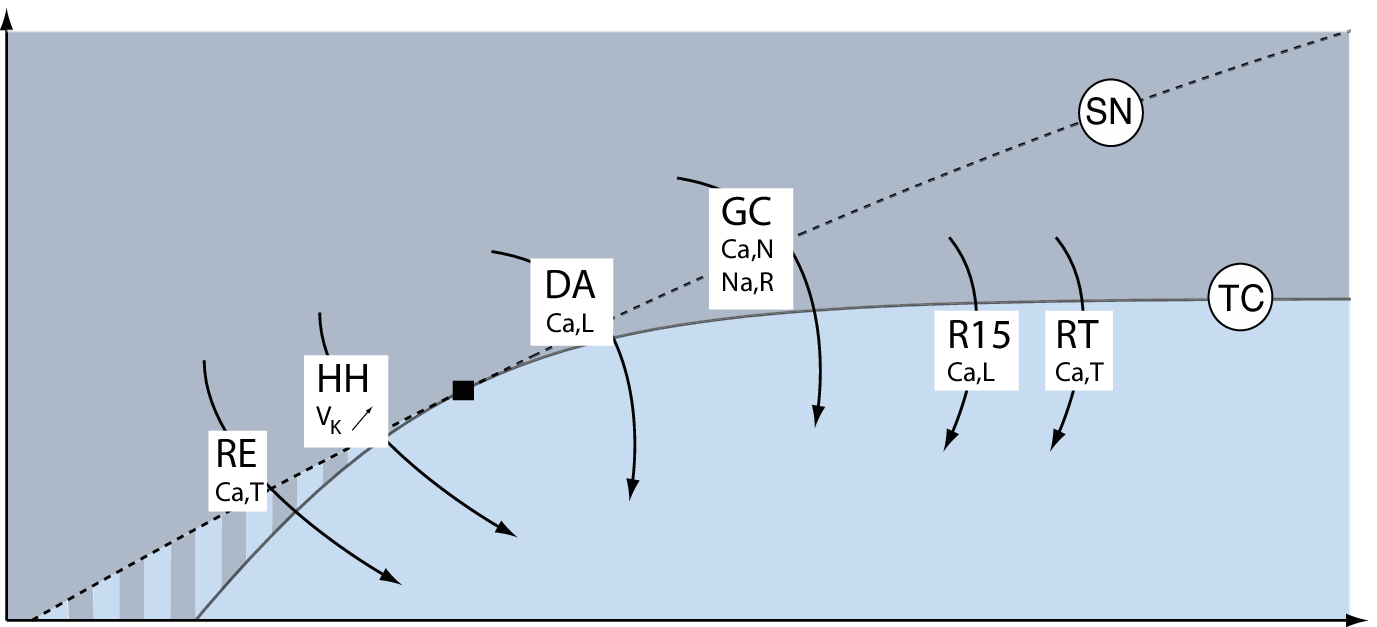}
\end{center}
\caption{{\bf Modifications in the balance between {restorative} and {regenerative} channels induce excitability switches in conductance-based models.} The figure sketches excitability switches of the Hodgkin-Huxley (HH) model \cite{main}{HODHUX}, Aplysia's R15 neuron (R15) model \cite{main}{rinzel1987dissection}, a dopaminergic (DA) neuron model \cite{main}{DRMASESE11}, thalamic reticular (RT) and relay (RE) neuron models \cite{main}{destexhe1998dendritic,destexhe1996vivo}, and a cerebral granule cell (GC) model \cite{main}{d2001theta} on the excitability parameter map computed for the two-dimensional model of \cite{main}{FRDRSE_SIAM}. All these conductance-based models can switch between {restorative} and {regenerative} excitability through the physiologically relevant regulation of specific ion channels.}
\label{FIG:CCfin}
\end{figure}

We now apply this algorithm to a number of published conductance-based models and show that all these models can switch between {restorative} and {regenerative} excitability through a transcritical bifurcation, as sketched in Fig. \ref{FIG:CCfin}. Figure \ref{FIG:CCfin} indicates two qualitatively distinct paths from {restorative} to {regenerative} excitability: {one path traversing the {mixed} excitability} region just described with Hodgkin-Huxley model (Fig. \ref{FIG: TC in HH}A) and {one path switching directly from restorative to regenerative excitability through the TC bifurcation} that will be illustrated on the dopaminergic neuron model.


\vspace{3mm}
\noindent\underline{\it\large Dopaminergic (DA) neuron model}\\

Model equations and parameters are taken from \cite{main}{DRMASESE11}. This is the model that  originally motivated \cite{main}{DRFRSESE_PLoS_sub}.

\newcounter{DA}
\begin{enumerate}[(i)]
\item {\bf Classification of gating variables as fast ($\mathcal G_{F}$), {slow} ($\mathcal G_{S}$), and adaptation ($\mathcal G_{A}$) variables}\\
The model includes fast sodium channels ($I_{Na,f}$), delayed-rectifier potassium channels ($I_{K,DR}$), L-type calcium channels ($I_{Ca,L}$), small conductance calcium-activated potassium (SK) channels ($I_{K,Ca}$), and calcium pumps ($I_{Ca,pump}$). We classify model variables as follows
\begin{itemize}
\item $\mathcal G_{F}=\{m_{Na,f}\}$
\item $\mathcal G_{S,-}=\{h_{Na,f},\,m_{K,DR}\}$ and $\mathcal G_{S,+}=\{m_{Ca,L}\}$
\item $\mathcal G_A=\{[Ca^{2+}]_{in}\}$
\end{itemize}
In order to unfold excitability switches, SK channel density is set to zero, since SK channels drastically attenuate DA neuron excitability by activating a strong calcium regulated potassium current \cite{main}{waroux2005sk,drion2012mitochondrion,ji2009tuning}. The intracellular calcium concentration is fixed at $[Ca^{2+}]_{in}=300\,nM$.

\item {\bf Balance equation and choice of the bifurcation parameter}\\
The only source of {regenerative} excitability is provided by L-type calcium channels. The balance equation reads
\begin{IEEEeqnarray}{rCl}\label{EQ: balance DA neurons}
\left.\frac{\partial\dot V}{\partial m_{Ca,L}}\frac{\partial m_{Ca,L,\infty}}{\partial V}\right|_{TC}=-\left.\left( \frac{\partial\dot V}{\partial h_{Na,f}}\frac{\partial h_{Na,f,\infty}}{\partial V}+\frac{\partial\dot V}{\partial m_{K,DR}}\frac{\partial m_{K,DR,\infty}}{\partial V}\right)\right|_{TC}.
\end{IEEEeqnarray}
We use the L-type calcium channel density $\bar g_{Ca,L}$ as the bifurcation parameter ({\it i.e.} $\lambda=\bar g_{Ca,L}$).
\end{enumerate}

\begin{figure}[h!]
\begin{center}
\includegraphics[width=0.9\linewidth]{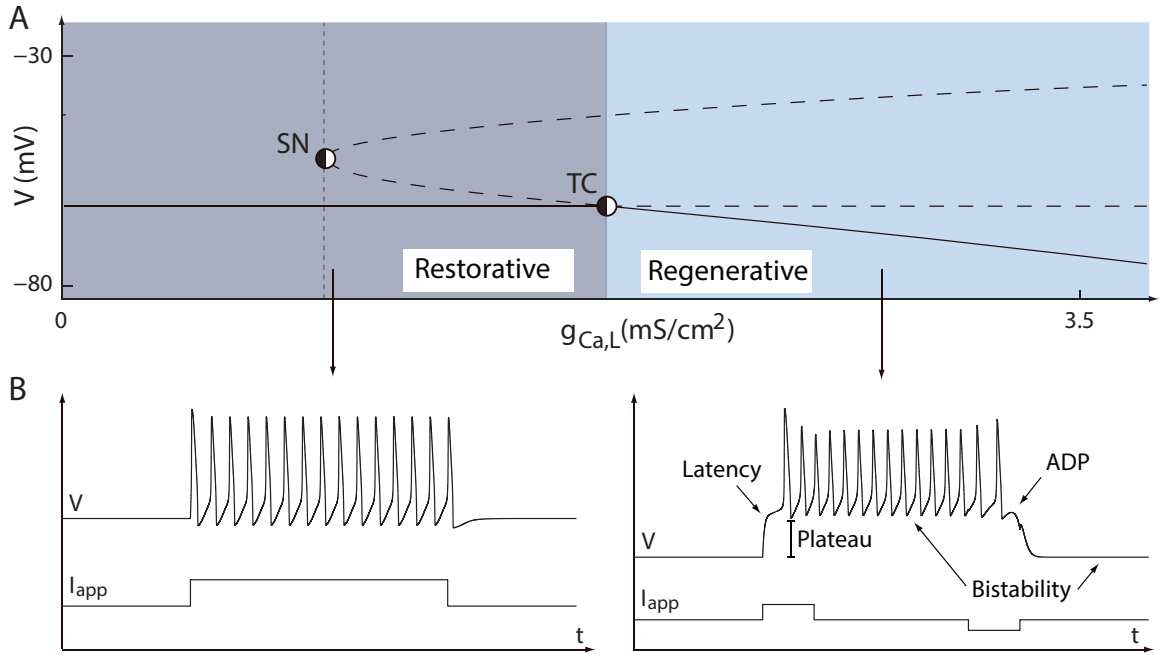}
\end{center}
\caption{{\bf Variations of L-type calcium channel density $\bar g_{Ca,L}$ induce excitability switches in a model of DA neurons \cite{main}{DRMASESE11}}. {\bf A.} Bifurcation diagram of the model with $\bar g_{Ca,L}$ as the bifurcation parameter. TC denotes a transcritical bifurcation, SN a saddle-node bifurcation. Branches of stable fixed points are represented as solid line, branches of saddle points and unstable points as dashed lines. {\bf B.} Electrophysiological responses of the model to step inputs of excitatory/inhibitory current (the intracellular calcium concentration is fixed at $[Ca^{2+}]_{in}=300\,nM$, which is within the physiological range). For $\bar g_{Ca,L}$ lower (resp. higher) than the critical value $\bar g_{Ca,L}^c$, the model exhibits typical electrophysiological signature of {restorative} (resp. {regenerative}) excitability. The low $\bar g_{Ca,L}$ configuration corresponds to $\bar g_{Ca,L}=1mS/cm^2$, whereas the high $\bar g_{Ca,L}$ configuration corresponds to $\bar g_{Ca,L}=3mS/cm^2$.}
\label{FIG:DAbif}
\end{figure}

\noindent {Solving Steps (iii) and (iv) of the algorithm above gives the following results (in this and the subsequent tables, the reported critical values are not exact, but rounded to the last significant digit)}:\\
\renewcommand{\arraystretch}{2}
\begin{center}
\begin{tabular}{c|c|c}
\newline
$V^c$ & $\bar g_{Ca,L}^c$ & $I^c$\\
\hline
$-64.9167$ mV & $1.9473$ mS/cm$^2$ & $9.5436$ $\mu$A/cm$^2$\\
\end{tabular}
\end{center}
{The} critical value $\bar g_{Ca,L}^c$ is roughly 1.5 times smaller than the nominal parameter value in \cite{main}{DRMASESE11}, which is consistent with the observation that the original model exhibits {regenerative} excitability during SK channel blockade.\\
%
The resulting bifurcation diagram is drawn in Fig. \ref{FIG:DAbif}A. In addition to confirming the existence of a transcritical bifurcation for the computed values, it reveals the excitability switches induced by changes in L-type calcium channel density in this model: in the absence of L-type calcium channels, the model exhibits {restorative} excitability. As $\bar g_{Ca,L}$ increases, a saddle point and an unstable node emerge at a saddle-node bifurcation, {which induces no excitability switch}. Further increase of $\bar g_{Ca,L}$ causes a transcritical bifurcation, where the stable point and the saddle exchange their stability. At this point, the stable steady-state becomes {regenerative}, and the model switches to {regenerative} excitability.

These excitability switches induce the predicted changes in the electrophysiological signatures, as illustrated in Fig. \ref{FIG:DAbif}B. Whereas the DA neuron model instantaneously reacts to a step input of depolarizing current for $\bar g_{Ca,L} < \bar g_{Ca,L}^c$, it exhibits electrophysiological signature of {regenerative} excitability such as spike latency, plateau oscillations and ADP as soon as $\bar g_{Ca,L}$ becomes higher than $\bar g_{Ca,L}^c$. In addition, the model becomes strongly bistable. 

\vspace{3mm}
\noindent\underline{\it\large Thalamic relay (RE) neuron model}\\

Model equations and parameters are taken from \cite{main}{destexhe1998dendritic}.

\begin{enumerate}[(i)]
\item {\bf Classification of gating variables as fast ($\mathcal G_{F}$), {slow} ($\mathcal G_{S}$), and adaptation ($\mathcal G_{A}$) variables}\\
The model includes fast sodium channels $I_{Na,f}$, delayed-rectifier potassium channels $I_{K,DR}$ and T-type calcium channels $I_{Ca,T}$. We classify model variables as follows

\begin{itemize}
\item $\mathcal G_{F}=\{m_{Na,f}\}$
\item $\mathcal G_{S,-}=\{h_{Na,f},\,m_{K,DR}\}$ and $\mathcal G_{S,+}=\{m_{Ca,T}\}$
\item $\mathcal G_A=\{h_{Ca,T}\}$
\end{itemize}

\item {\bf Balance equation and choice of the bifurcation parameter}\\
The only source of {regenerative} excitability is provided by T-type calcium channels. {The associated balance equation reads}
\begin{IEEEeqnarray*}{rCl}
\left.\frac{\partial\dot V}{\partial m_{Ca,T}}\frac{\partial m_{Ca,T,\infty}}{\partial V}\right|_{TC}=-\left.\left( \frac{\partial\dot V}{\partial h_{Na,f}}\frac{\partial h_{Na,f,\infty}}{\partial V}+\frac{\partial\dot V}{\partial m_{K,DR}}\frac{\partial m_{K,DR,\infty}}{\partial V}\right)\right|_{TC}.
\end{IEEEeqnarray*}
{T-type calcium channels are} dynamically regulated by a (slow) {voltage-gated inactivation} $h_{Ca,T}$. We use this {variable} as the bifurcation parameter ({\it i.e.} $\lambda=h_{Ca,T}$).
\end{enumerate}

\begin{figure}[h!]
\begin{center}
\includegraphics[width=0.9\linewidth]{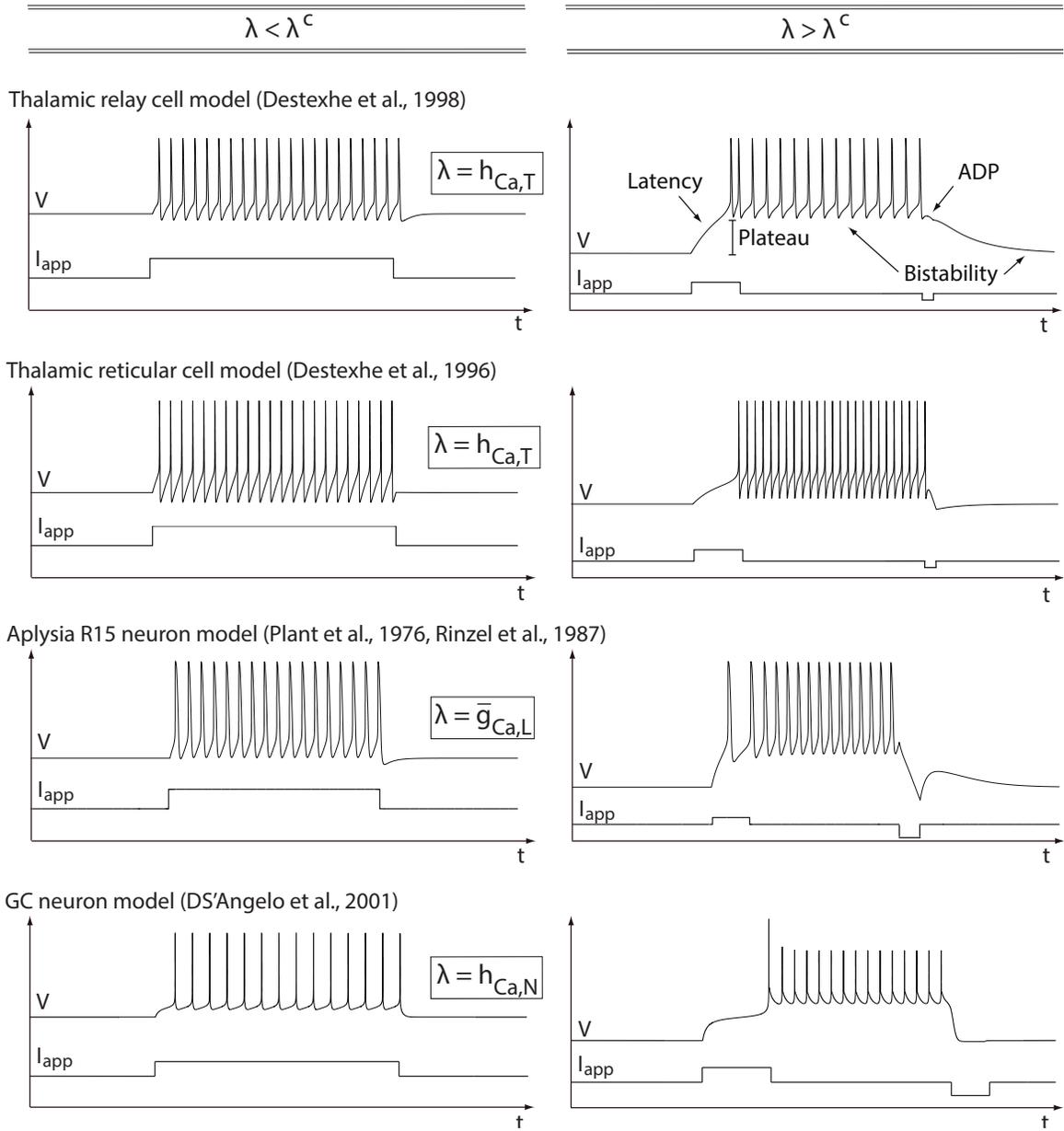}
\end{center}
\caption{{\bf The same mathematical bifurcation in different conductance-base models causes the same switch in electrophysiological signatures.} The figure shows the electrophysiological responses of various conductance-based models to step inputs of excitatory/inhibitory current when the bifurcation parameter $\lambda$ is lower (left) or higher (right) than the critical value $\lambda^c$. This bifurcation parameter can be either the density or the inactivation variable of a {regenerative} channel. Other adaptation variables are set to constant values chosen in physiological ranges (see text). For $\lambda$  lower (resp. higher) than the critical value $\lambda^c$, all models exhibit electrophysiological signatures of {restorative} (resp. {regenerative}) excitability. Numerical values of the parameter $\lambda$ in the different plots are as follows. Thalamic relay cell: left $h_{Ca,T}=0$, right $h_{Ca,T}=0.2$. Thalamic reticular cell: left $h_{Ca,T}=0$, right $h_{Ca,T}=0.4$. Aplysia R15 neuron: left $\bar g_{Ca,L}=10^{-6}mS/cm^2$, right $\bar g_{Ca,L}=0.004mS/cm^2$. GC neuron: left $h_{Ca,N}=0.01,\,h_{Na,R}=0.1$, right $h_{Ca,N}=0.3,\,h_{Na,R}=0.1$.}
\label{FIG:switches}
\end{figure}

\noindent {Solving Steps (iii) and (iv) of the algorithm above gives the following results}:\\
\renewcommand{\arraystretch}{2}
\begin{center}
\begin{tabular}{c|c|c}
\newline
$V^c$ & $\bar h_{Ca,T}^c$ & $I^c$\\
\hline
$-61.5561$ mV & $0.0041$ & $0.9074$ nA/cm$^2$\\
\end{tabular}
\end{center}
{Since $\bar h_{Ca,T}^c\in(0,\,1)$, the model dynamically switches between restorative and regenerative excitability when $h_{Ca,T}$ crosses the critical value,} and the electrophysiological signatures are consistent with the excitability switches. See Fig. \ref{FIG:switches}.

\vspace{3mm}
\noindent\underline{\it\large Thalamic reticular (RT) neuron model}\\

Model equations and parameters are taken from \cite{main}{destexhe1996vivo}, maximal conductances are adapted as in \cite{main}{DRFRSESE_PLoS_sub}.

\begin{enumerate}[(i)]
\item {\bf Classification of gating variables as fast ($\mathcal G_{F}$), {slow} ($\mathcal G_{S}$), and adaptation ($\mathcal G_{A}$) variables}\\
The model includes fast sodium channels $I_{Na,f}$, delayed-rectifier potassium channels $I_{K,DR}$ and T-type calcium channels $I_{Ca,T}$. We classify model variables as follows

\begin{itemize}
\item $\mathcal G_{F}=\{m_{Na,f}\}$
\item $\mathcal G_{S,-}=\{h_{Na,f},\,m_{K,DR}\}$ and $\mathcal G_{S,+}=\{m_{Ca,T}\}$
\item $\mathcal G_A=\{h_{Ca,T}\}$
\end{itemize}

\item {\bf Balance equation and choice of the bifurcation parameter}\\
The only source of {regenerative} excitability is provided T-type calcium channels. {The associated balance equation has the same structure as for the thalamic relay neuron model considered above. Along the same line, we choose the T-type calcium channel inactivation $h_{Ca,T}$ as the bifurcation parameter ({\it i.e.} $\lambda=h_T$).}

\end{enumerate}

\noindent {Solving Steps (iii) and (iv) of the algorithm above gives the following results}:\\
\renewcommand{\arraystretch}{2}
\begin{center}
\begin{tabular}{c|c|c}
\newline
$V^c$ & $\bar h_{Ca,T}^c$ & $I^c$\\
\hline
$-48.8063$ mV & $0.1780$ & $-0.82484$ nA\\
\end{tabular}
\end{center}
As in the case of the thalamic relay neuron model, the T-type calcium channel inactivation generates a dynamical switch between {restorative} and {regenerative} excitability, significantly affecting neuron response to external inputs (Fig. \ref{FIG:switches}).

\vspace{3mm}
\noindent\underline{\it\large Aplysia R15 neuron model}\\

Model equations and parameters are taken from \cite{main}{rinzel1987dissection}.

\begin{enumerate}[(i)]
\item {\bf Classification of gating variables as fast ($\mathcal G_{F}$), {slow} ($\mathcal G_{S}$), and adaptation ($\mathcal G_{A}$) variables}\\
The model includes fast sodium channels $I_{Na,f}$, delayed-rectifier potassium channels $I_{K,DR}$, slow L-type calcium channels $I_{Ca,L}$ and calcium-activated potassium channels $I_{K,Ca}$. We classify model variables as follows
\begin{itemize}
\item $\mathcal G_{F}=\{m_{Na,f}\}$
\item $\mathcal G_{S,-}=\{h_{Na,f},\,m_{K,DR}\}$ and $\mathcal G_{S,+}=\{m_{Ca,L}\}$
\item $\mathcal G_A=\{[Ca^{2+}]_{in}\}$
\end{itemize}
The intracellular calcium conductance is fixed at $[Ca^{2+}]_{in}=0.09nM$.

\item {\bf Balance equation and choice of the bifurcation parameter}\\
As in the case of DA neurons, the source of {regenerative} excitability is provided by L-type calcium channels, and we take their maximal conductance $\bar g_{Ca,L}$ as the bifurcation parameter. The associated balance equation has the same structure as for the DA neuron model considered above.
\end{enumerate}

\noindent {Solving Steps (iii) and (iv) of the algorithm above gives the following results}:\\
\renewcommand{\arraystretch}{2}
\begin{center}
\begin{tabular}{c|c|c}
\newline
$V^c$ & $\bar g_{Ca,L}^c$ & $I^c$\\
\hline
$-48.0516$ mV & $5.4054\mbox{ }10^{-5}$ mS/cm$^2$ & $-0.0318$ $\mu$A/cm$^2$\\
\end{tabular}
\end{center}
Comparing the critical value $\bar g_{Ca,L}^c$ with the original value $\bar g_{Ca,L} = 4\mbox{ }10^{-3}$ mS/cm$^2$ shows that the bursting model proposed in \cite{main}{rinzel1987dissection} exhibits strong {regenerative} excitability. Switches of electrophysiological signatures are illustrated in Fig. \ref{FIG:switches}.

\vspace{3mm}
\noindent\underline{\it\large Cerebellar granular cell (GC) model}\\

Model equations and parameters are taken from \cite{main}{d2001theta}.
\begin{enumerate}[(i)]
\item {\bf Classification of gating variables as fast ($\mathcal G_{F}$), {slow} ($\mathcal G_{S}$), and adaptation ($\mathcal G_{A}$) variables}\\
The model includes the following ion channels: fast ($I_{Na,f}$), persistent ($I_{Na,P}$) and resurgent sodium channels ($I_{Na,R}$) ; N-type calcium channels ($I_{Ca,N}$) ; delayed rectifier ($I_{K,DR)}$, A-type ($I_{K,A}$), inward rectifier ($I_{K,IR}$), calcium activated ($I_{K,Ca}$) and slow potassium channels ($I_{K,slow}$). We classify model variables as follows
\begin{itemize}
\item $\mathcal G_{F}=\{m_{Na,f},\,m_{Na,P}\}$
\item $\mathcal G_{S,-}=\{h_{Na,f},\,m_{K,DR},\,m_{K,A},\,m_{K,IR}\}$ and $\mathcal G_{S,+}=\{m_{Na,R},\,m_{Ca,N}\}$
\item $\mathcal G_A=\{h_{Na,R},\,h_{Ca,N},\,h_{K,A},\,m_{K,slow}\}$
\end{itemize}
We set the persistent and calcium-activated currents to zero (these two channels do not impact excitability type as anticipated by our classification and shown by D'Angelo and colleagues \cite{main}{d2001theta}). The inactivation of the A-type potassium current is fixed at $h_{K,A} = 0.02$ and the activation of the slow potassium current is fixed at $m_{K,slow} = 0.13$.

\item {\bf Balance equation and choice of the bifurcation parameter}\\
The neuron model possesses two sources of {regenerative} excitability: resurgent sodium channels and N-type calcium channels. As in the case of T-type calcium channels mentioned above, these two channels possess an inactivation gate, which is used as the bifurcation parameter. {We apply our algorithm by varying the parameter of one regenerative current while fixing the other at different values. This permits to draw an approximated hypersurface in the $(h_{Ca,N},h_{Na,R})$ plane at which the balance equation is satisfied and the model undergoes the transcritical bifurcation and the associated excitability switch}. {The two balance equations read, respectively}:
\begin{IEEEeqnarray*}{rCl}
\left.\frac{\partial\dot V}{\partial m_{Na,R}}\frac{\partial m_{Na,R,\infty}}{\partial V}\right|_{TC}&=&-\left.\left( \frac{\partial\dot V}{\partial h_{Na,f}}\frac{\partial h_{Na,f,\infty}}{\partial V}+\mathlarger{\mathlarger{\sum}}_{y=DR,\,A,\,IR}\frac{\partial\dot V}{\partial m_{K,y}}\frac{\partial m_{K,y,\infty}}{\partial V}+\frac{\partial\dot V}{\partial m_{Ca,N}}\frac{\partial m_{Ca,N,\infty}}{\partial V}\right)\right|_{TC},\\
\left.\frac{\partial\dot V}{\partial m_{Ca,N}}\frac{\partial m_{Ca,N,\infty}}{\partial V}\right|_{TC}&=&-\left.\left( \frac{\partial\dot V}{\partial h_{Na,f}}\frac{\partial h_{Na,f,\infty}}{\partial V}+\mathlarger{\mathlarger{\sum}}_{y=DR,\,A,\,IR}\frac{\partial\dot V}{\partial m_{K,y}}\frac{\partial m_{K,y,\infty}}{\partial V}+\frac{\partial\dot V}{\partial m_{Na,R}}\frac{\partial m_{Na,R,\infty}}{\partial V}\right)\right|_{TC},
\end{IEEEeqnarray*}
\end{enumerate}

\begin{figure}[h!]
\begin{center}
\includegraphics[width=0.9\linewidth]{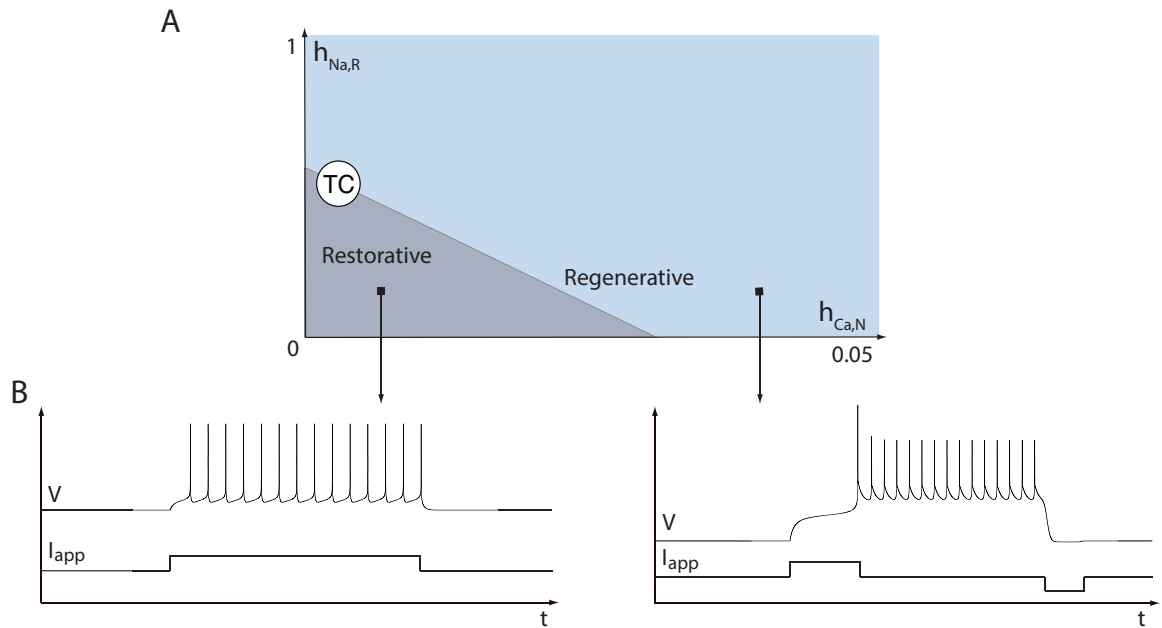}
\end{center}
\caption{{\bf Joint variations of the inactivation gates of $N$-type calcium channels and resurgent sodium channels induce excitability switches in cerebellar granular cells.} {\bf A.} Two parameter bifurcation diagram of the mode with $h_{Ca,N}$ and $h_{Na,R}$ as bifurcation parameters. TC denotes a branch of transcritical bifurcations detected following the algorithm in Table \ref{TAB: TC algo}. {\bf B.} Electrophysiological responses of the model to step inputs of excitatory/inhibitory current: left $h_{Ca,N}=0.01,\,h_{Na,R}=0.1$, right $h_{Ca,N}=0.3,\,h_{Na,R}=0.1$.}
\label{FIG:GCbif}
\end{figure}

\noindent {Solving Steps (iii) and (iv) of the algorithm one obtains the parameter chart in Fig.} \ref{FIG:GCbif}. {These results show that both channels can induce a dynamical switch in excitability. However, the N-type calcium channel contributes much more to regenerative excitability than the resurgent sodium channel in this model: as soon as $h_{Ca,N}\gtrsim 0.03$ the model is in regenerative excitability for all values of $h_{Na,R}$. On the contrary, when $h_{Ca,N}=0$ the inactivation of resurgent sodium channel should be more than a half de-inactivated for the model to exhibit regenerative excitability.}

\subsection*{The {balance equation} determines a switch from restorative to regenerative excitability}

{As illustrated in Figure} \ref{FIG: SIAM PC}, {the significance of the transcritical bifurcation is
that it delineates in the parameter space the boundary of a specific type of  excitability
and that this boundary is determined by a simple physiological balance (Eq. \ref{EQ: fundamental}) between restorative
and regenerative channels.}

{Specific to regenerative  excitability is the bistable (right) phase portrait of Figure} \ref{FIG: SIAM exc types}.
{For the five analyzed conductance-based models, our bifurcation analysis of the full model
confirms the existence of a bistable range beyond the transcritical bifurcation, where a regenerative
resting state and a spiking limit cycle coexist. In each case, the bistability range is obtained for the nominal time
scales of the published model and is robust to a variation of time scales. In each case, the bistabilty range
is also neuromodulated, that is,  determined by conductance parameters that are known to vary in slower time scales
and/or across neurons of a same type.}

\begin{figure}[h!]
\begin{center}
\includegraphics[width=0.7\linewidth]{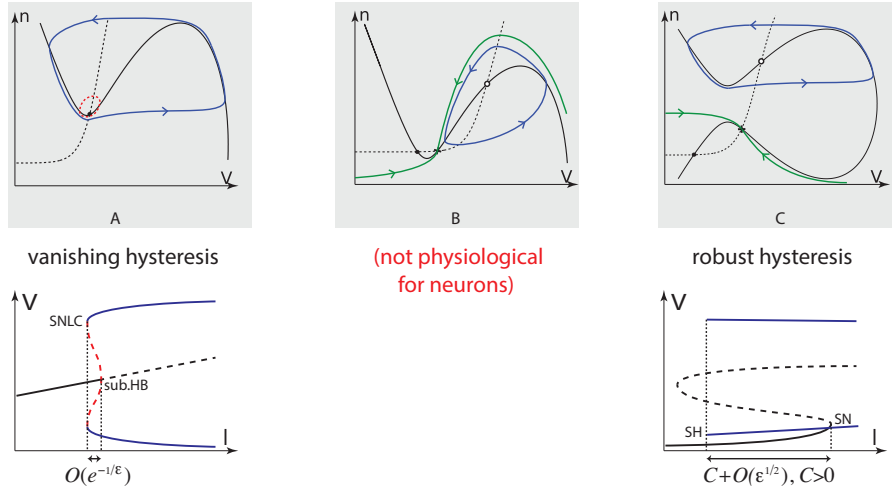}
\end{center}
\caption{{\bf Three bistable phase portraits of model (\ref{EQ: SIAM planar model}) and cartoon of the associated hysteretic bifurcation diagrams.} In the phase portraits, a solid line denotes the $V$-nullcline, whereas a dashed line denotes the $n$-nullcline. Stable fixed points are depicted as filled circles, whereas unstable as circles and saddle points as cross. Stable limit cycles are drawn as solid oriented blue curves, whereas unstable as red dashed curves. The stable manifolds of saddle points are depicted as green oriented curves. In bifurcation diagrams, a solid line denotes branches of stable fixed points, whereas a dashed line denotes branches of unstable or saddle points. Branches of stable limit cycles are depicted as blue lines, whereas branches of unstable limit cycles as red dashed lines. {\it sub.HB} denotes a subcritical Hopf bifurcation, {\it SNLC} a saddle-node limit cycles bifurcation, {\it SN} a saddle-node bifurcation, and {\it SH} a saddle-homoclinic bifurcation.\\
\quad{\bf A-B.} Restorative bistability. {\bf A.} Subcritical Hopf bifurcation. Hysteresis vanishes exponentially fast as timescale separation increases. {\bf B.} Restorative saddle-homoclinic bifurcation. Not physiological because it violates the time scale separation between $V$ and $n$.\\\quad {\bf C.} Regenerative bistability ruled by a regenerative saddle-homoclinic bifurcation. Hysteresis is barely affected by time-scale separation.}
\label{FIG:bistability types}
\end{figure}

{It is important to distinguish this robust and physiologically regulated bistability from other types of bistability that can be encountered in conductance-based
models.   Figure} \ref{FIG:bistability types}  {qualitatively illustrates three typical bistable phase portraits associated to the planar model} (\ref{EQ: SIAM planar model}){ that exhibit the coexistence of a stable resting state and of a spiking limit cycle. The first two are associated to restorative excitability and are extensively studied in the literature. See, e.g.,} \cite{main}{Rinzel:1989:ANE:94605.94613,IZHIKEVICH2007}{ and references therein. Only the third one is associated to regenerative
excitability.}

{The three bistable phase portraits share the common feature of ``hard excitation": as the amplitude of a step input depolarizing current
is increased, the response of the neuron abruptly switches from no oscillation to high frequency spiking. Following the historical
classification of Hodgkin \cite{main}{hodgkin1948local}, the three situations correspond to Class II neurons, as opposed to Class I neurons for which the spiking frequency
gradually increases with the depolarizing current amplitude.}

{Hard excitation can be a manifestation either of a switch-like monostable bifurcation diagram or of a hysteretic bistable bifurcation
diagram. By definition, the three bistable phase portraits in Figure} \ref{FIG:bistability types} {give rise to  hysteretic bifurcation diagrams.
But for the two bistable phase portraits associated to restorative excitability, the hysteresis is highly dependent on the time scale separation, {\it i.e.}, {the ratio $\varepsilon$ between the fast and slow time constants}. In the case of  the first phase portrait (subcritical Hopf bifurcation), asymptotic analysis shows that the hysteresis vanishes as $\mathcal O(e^{-1/\varepsilon})$. In the case of
second  phase portrait (saddle-homoclinic bifurcation), the situation is even worse because for small $\epsilon>0$ the system necessarily undergoes a monostable saddle-node on invariant circle bifurcation. In fact, the second phase portrait is not physiological for conductance based-models.  For instance, in the hypothetical $I_{Na,p} + I_{K}$ conductance-based model considered in} \cite[Fig. 6.44]{main}{IZHIKEVICH2007},{ the time constant of the potassium
activation must be set to below $0.17ms$ to create a saddle-homoclinic  bifurcation, which is roughly 40 times smaller than its
physiological value and even smaller that the fast time constant. A geometric proof of the generality of this fact is provided in} \cite{main}{FRDRSE_SIAM}. {The conclusion is that hysteresis associated to restorative excitability is at best very small (if any) in physiologically plausible conductance based models, which makes their electrophysiological signatures similar to those associated to a switch-like monostable bifurcation diagram.}

{In sharp contrast, the hysteresis associated to regenerative excitability is barely affected by the time-scale separation. Instead
it is regulated by conductance parameters whose modulation is physiological (for instance, a regenerative ion channel density).
The extended hysteresis is what determines the specific electrophysiological signature of regenerative excitability: a pronounced
spike latency, a possible plateau oscillation, and an after depolarization potential. As a consequence, those features cannot be
robustly reproduced in physiologically plausible conductance based models of  restorative excitability. Because those features are
important markers of modern electrophysiology }\cite{main}{Fuentealba2005,Junek2010},{ the distinction between restorative and regenerative
excitability seems physiologically relevant, beyond the possible shared feature of hard excitation.}

{In conclusion, the  bistability associated to regenerative excitability is specific in that it produces a robust
electrophysiological signature in physiologically plausible parameter ranges and
consistent with many experimental observations.  It is in that sense that {balance equation} delineates
a switch of excitability of physiological relevance.}

\subsection*{{A refined classification of neuronal excitability}}

Early in the history of neurodynamics \cite{main}{hodgkin1948local}, Hodgkin proposed a classification of excitability in three different classes: \\

 \underline{\it Class I}: The spiking frequency vs. input current amplitude (f/I) curve is continuous, {\it i.e.}, the spiking frequency continuously increases from zero to high-frequency firing as the input current amplitude rises. Class I excitability is also referred to as ``soft" excitation.\\

\underline{\it Class II}: The f/I is discontinuous, {\it i.e.}, the spiking frequency abruptly switches from zero to high-frequency firing as the amplitude of the applied current is raised above a certain threshold. Class II excitability is also referred to as ``hard" excitation.\\

\underline{\it Class III}: The spiking frequency is zero for all amplitudes of the applied current. Transient action potentials can be generated in response to high-frequency stimuli.\\

Because regenerative excitability exhibits hard excitation, it is a physiologically distinct subtype of Class II excitability.

\begin{figure}[h!]
\begin{center}
	\includegraphics[width=0.7\linewidth]{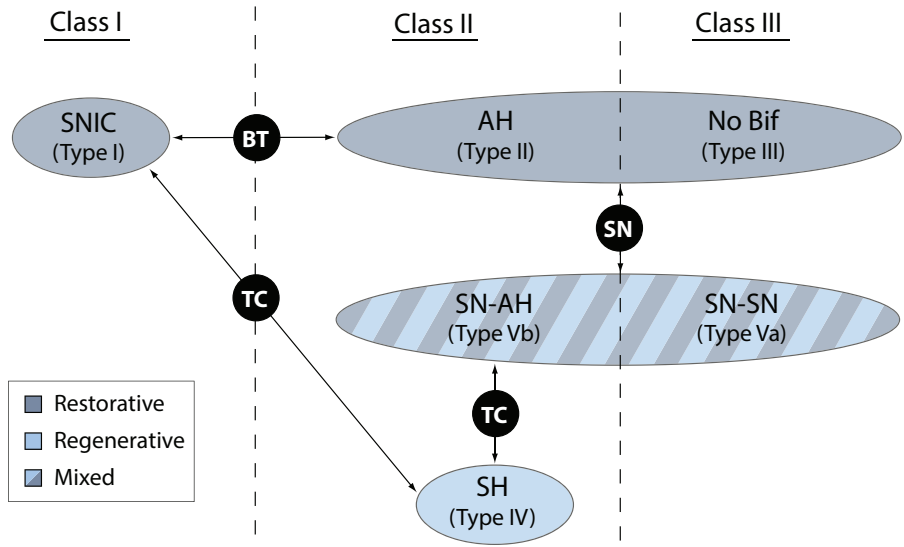}
\end{center}
\caption{{\bf The various bifurcations associated to different types of neuronal excitability.}  $SNIC$: saddle-node on invariant circle; $BT$: Bogdanov-Takens; $AH$: Andronov-Hopf; $SN$: saddle-node; $TC$: transcritical; $SH$: saddle-homoclinic. See also \cite{main}{FRDRSE_SIAM} for more detailed definitions and properties of excitability Types I-V and associated transition bifurcations in a planar neuron model.\\
\quad\underline{Class I} excitability occurs in the neighborhood of a SNIC bifurcation \cite{main}{ermentrout1996type} and is purely restorative.\\
\quad\underline{Class II} excitability can be either restorative in which case the stable equilibrium looses stability in  a subcritical Hopf bifurcation (Type II in \cite{main}{ermentrout1996type}) or regenerative in which case a stable equilibrium coexists with a stable limit cycle over a robust bistable range organized by a (singularly perturbed) saddle homoclinic bifurcation (Type IV in \cite{main}{FRDRSE_SIAM}). In a small parameter range, class II excitability can also exhibit a mixed type (Type Vb in \cite{main}{FRDRSE_SIAM}), where a regenerative "down" stable equilibrium  coexists with a "up" restorative stable equilibrium or limit cycle. Stability of those attractors is lost either in saddle-node or Hopf bifurcations.\\
\quad\underline{Class III} excitability can be either restorative (a monostable equilibrium) or exhibits a mixed type (Type Va in \cite{main}{FRDRSE_SIAM}), where a regenerative down stable equilibrium coexists with a restorative up stable equilibrium. Both attractors loose stability in a saddle-node bifurcation.\\
\quad The transition to regenerative excitability is always through a transcritical bifurcation.
%
}
\label{FIG:Hodgkin vs rest-reg}
\end{figure}

Bifurcation theory helps relating this physiological classification to mathematical signatures of the associated neuron models. Distinct bifurcations delineate the different excitability classes as well as the different excitability mechanisms within a given class. They are summarized in Fig. \ref{FIG:Hodgkin vs rest-reg}.

Ermentrout \cite{main}{ermentrout1996type} showed that Class I excitability arises from a saddle-node on invariant circle bifurcation, whereas Class II excitability arises from a Hopf bifurcation. Both bifurcations correspond to examples of restorative excitability in the terminology of the present paper and the transition between Class I and II is governed by a Bogdanov-Takens bifurcation.

Our recent paper \cite{main}{FRDRSE_SIAM} further expands this classification to account for regenerative excitability. Regenerative excitability (called Type IV in \cite{main}{FRDRSE_SIAM}) arises from a (singularly perturbed) saddle-homoclinic bifurcation and the transition from restorative to regenerative excitability always involves a transcritical bifurcation.

\section*{Discussion}

\subsection*{A simple and robust balance equation identifies a transcritical bifurcation in arbitrary conductance based models}
Motivated by a geometric analysis of a qualitative phase portrait, we have proposed an algorithm that {easily} detects a transcritical bifurcation in arbitrary conductance based models. Owing to the special structure of such models, the algorithm leads to solving an algebraic equation of remarkable simplicity and physiological relevance: a balance between {slow restorative} and {slow regenerative} ion channels. The condition is also robust because the balance is independent of the detailed kinetics, {even though it critically} relies on a classification of variables in three well separated time-scales, in {good agreement} with what is known on ion channels kinetics \cite{main}{HILLE84}.

\subsection*{The ubiquity of a transcritical bifurcation  in conductance-based models}

{The detection of the transcritical bifurcation relies on the sole existence of
a physiological balance between restorative and regenerative ion channels.
Given that all neuronal models possess restorative sodium and potassium channels,
this implies that a transcritical bifurcation exists in every conductance-based model
that possesses at least one regenerative ion channel. Moreover, the channel balance, and therefore the TC
bifurcation, are readily detectable in a  model of arbitrary dimension (both in the state and parameters): the balance} (\ref{EQ: fundamental}) {simply defines a hypersurface in the
parameter space that can algebraically be tracked under arbitrary parameter variations, as illustrated with in
the GC model above.}

{In spite of its ubiquity and of its physiological significance, we
are not aware of an earlier reference to a transcritical bifurcation in conductance based models.
A reason for this omission might be accidental:  there are no regenerative channels in the
 seminal model of Hodgkin and Huxley (unless one modifies the potassium resting potential $V_k$)
and  this model has been the inspiration of most
 mathematical analyses of conductance-based models.}

{For the same reason, it seems physiologically relevant to distinguish between restorative and regenerative excitability beyond Hodgkin's classification of {Class I (``soft'') and Class II (``hard") excitability}. Regenerative {(and restorative)} excitability faithfully capture the presence {(or the absence)} of specific electrophysiological signatures of modern electrophysiology such as spike latency, afterdepolarization potentials, or robust coexistence of  resting state and repetitive spikes.}
 
\subsection*{A {single} mathematical prediction applies to many distinct physiological observations}
 
Although purely mathematical in nature, the transcritical bifurcation has a remarkable predictive value in several published conductance based models. In each of the six analysed models, the proposed algorithm identifies a physiological parameter that acts as a tuner of neuronal excitability in a physiologically plausible range and in full agreement with existing experimental data. At the same time, the distinct nature of the regulating parameter, which can be either the maximal conductance or the inactivation gating variable of a {regenerative} ion channel depending on the neuron model, is associated to distinctly different regulation mechanisms.

\subsection*{Reduced modelling should retain the balancing channel}

The classification of gating variables in three distinct time scales is an essential modelling step both for the proposed algorithm and for the reduction of full conductance-based models to low-dimensional models that can be used in population studies \cite{main}{izhikevich2008large}. {Despite the inherent robustness of  time-scale separation analysis, this classification is a limitation of the proposed approach if the model contains ion channels with poorly known kinetics.}  When all {slow} ion channels are properly identified, they can be aggregated in a single {slow} variable to lead to a second order model of the type (\ref{EQ: SIAM planar model}), where the single parameter $n_0$ captures the {restorative} or {regenerative} nature of the aggregated {slow} variable. Further reduction to a one-dimensional hybrid model with reset is possible thanks to the time-scale sepration between the voltage $V$ and the {slow} variable $n$. This reduction is illustrated in \cite{main}{DRFRSESE_PLoS_sub} on the thalamic TC neuron and the reduced model remarkably retains the switch of excitability of its high-dimensional counterpart. In contrast, a reduced model will lose the switch of excitability of the full conductance-based model when a {regenerative} ion channel is treated as a fast gating variable. {It is for instance common in model reduction to set the activation a calcium channel to steady state. This amounts to treat the calcium activation as a fast variable, which makes the channel ``slow restorative" in the terminology of this paper. If the calcium channel is the only source of regenerative excitability, then the reduced model will not retain features of regenerative excitability}.

\subsection*{Neuronal excitability is regulated}

{In each of the analysed conductance-based models, the balance equation responsible for the switch of excitability
is satisfied for a set of parameters that is close to the published  parameter values. This observation supports the hypothesis
that neuronal excitability is tightly regulated by molecular mechanisms and that the influence
of  the channel  balance condition on neuronal excitability might play a role in
neuronal signaling.}

\section*{Materials and Methods}

\subsection*{Numerical analysis}

Numerical temporal traces of the different neurons (Figs. \ref{FIG: SIAM exc types}, \ref{FIG: TC in HH}, \ref{FIG:DAbif}, \ref{FIG:switches}) were reproduced by implementing in MATLAB ( available at {\tt http://www.mathworks.com}) the original models as described in the associated papers. The phase portraits in Figures \ref{FIG: SIAM exc types} and \ref{EQ: geom trans constr} were hand-drawn using the Open Source vector graphics editor Inkscape ({\tt http://inkscape.org}). The phase portraits in Figure \ref{FIG: TC in HH} were numerically drawn with MATLAB by implementing the planar model in \cite{main}{rinzel1985excitation} and subsequently modified with Inkscape. The bifurcation diagrams in Figures \ref{FIG: TC in HH}, \ref{FIG:CCfin}, and \ref{FIG:DAbif} were drawn by implementing the algorithm of Table \ref{TAB: TC algo} in MATLAB.  No figure or part of figure was reproduced from other published works.

\section*{Acknowledgment}
{The reviewers are gratefully acknowledged for insightful comments and suggestions about the original version of the manuscript.}

\bibliographystyle{main}{plos2009}
\bibliography{main}{../../../../refs}{References}

\newpage

\newpage

\begin{table}[htbp]
\begin{center}
\begin{tabular}{l|c|c}
\textbf{Ion Channel} & \textbf{Gating Kinetics} &\textbf{Classification}\\
\hline
\hline
{\color{white}.}& {\color{white}.}&{\color{white}.}\\
\textbf{Sodium Channels} &  & \\
{\color{white}.}& {\color{white}.}&{\color{white}.}\\
Transient & FA, SI - Negative feedback via SI& \colorbox{white}{\phantom{aaaaaaaa} {{Slow} restorative} \phantom{aaaaaaaa}} \\
Persistent & FA - No slow gating variable & Neutral\\
Resurgent & SA, USI - Positive feedback via SA & \colorbox{white}{\phantom{aaaaaaa.} {{Slow} regenerative} \phantom{aaaaaaa}}\\
\hline
{\color{white}.}& {\color{white}.}&{\color{white}.}\\
\textbf{Calcium Channels} &  & \\
{\color{white}.}& {\color{white}.}&{\color{white}.}\\
L-type &SA - Positive feedback via SA \ & \colorbox{white}{\phantom{aaaaaaa.} {{Slow} regenerative} \phantom{aaaaaaa}} \\
T-type & SA, USI - Positive feedback via SA & \colorbox{white}{\phantom{aaaaaaa.} {{Slow} regenerative} \phantom{aaaaaaa}} \\
N, P/Q, R-type & SA, USI - Positive feedback via SA & \colorbox{white}{\phantom{aaaaaaa.} {{Slow} regenerative} \phantom{aaaaaaa}} \\
\hline
{\color{white}.}& {\color{white}.}&{\color{white}.}\\
\textbf{Potassium Channels} &  & \\
{\color{white}.}& {\color{white}.}&{\color{white}.}\\
Delayed Rectifiers & SA - Negative feedback via SA& \colorbox{white}{\phantom{aaaaaaaa} {{Slow} restorative} \phantom{aaaaaaaa}}  \\
KCNQ &USA - No slow gating variable   & Neutral\\
eag/erg & USA - No slow gating variable &  Neutral\\
A-type & FA, SI - Positive feedback via SI& \colorbox{white}{\phantom{aaaaaaa.} {{Slow} regenerative} \phantom{aaaaaaa}} \\
BK & SA - Negative feedback via SA&  \colorbox{white}{\phantom{aaaaaaaa} {{Slow} restorative} \phantom{aaaaaaaa}} \\
HCN &  SA - Negative feedback via SA &\ \colorbox{white}{\phantom{aaaaaaaa} {{Slow} restorative} \phantom{aaaaaaaa}}
\vspace{3mm}
\end{tabular}\\
{\small FA: fast activation, FI: fast inactivation, SA: slow activation, SI: slow inactivation, USA: utraslow activation, USI: ultraslow inactivation}
\end{center} \caption{{\bf Classification of ion channels according to their gating kinetics.} Activation and inactivation variables are distributed in three groups: fast, {slow}, and ultra-slow (adaptation). {Slow} variables are defined as {restorative} (resp. {regenerative}) if they induce a negative (resp. positive) feedback on membrane potential variations. An ion channel that posses a {slow} {restorative} variable is called ``{slow} {restorative} channel'', and similarly for {slow} {regenerative} channels. Channels that do not posses a {slow} variable are called ``neutral channels''. This classification might change for a given channel for some channel subtypes.} \label{TAB:IC}
\end{table}

\begin{table}
\begin{center}
{\bf Algorithm for the detection of a transcritical bifurcation in generic conductance-based models via modulation of a regenerative ionic current and computation of the excitability switch bifurcation diagram.}
\vspace{-2mm}
\end{center}
\begin{tabular}{|l|}
\hline
\noindent(i) {\bf Classification of gating variables as fast ($\mathcal G_{F}$), {slow} ($\mathcal G_{S}$), and adaptation ($\mathcal G_{A}$) variables}\\
\hspace{5mm}(i-a) Following Tab. \ref{TAB:IC}, group gating variables in the three groups $\mathcal G_{F}$, $\mathcal G_{S}$, and $\mathcal G_{A}$.\\
\hspace{5mm}(i-b) Split $\mathcal G_{S}$ in {regenerative} $\mathcal G_{S,+}$ and {restorative} $\mathcal G_{S,-}$ {slow} gating variables.\\
\hspace{5mm}(i-c)  If adaptation variables are present, set them to constant {physiologically relevant values}.\\
\hline
\noindent(ii) {\bf Balance equation and choice of the bifurcation parameter}\\
\hspace{5mm}(ii-a) Select a {regenerative} ionic current $I_{reg}$ and the associated {regenerative} {slow} gating\\
\hspace{5mm}\hspace{7.5mm} variable $x^{reg}$.\\
\hspace{5mm}(ii-b) Write the balance equation\\
\hspace{0.25\textwidth}$\displaystyle{\left.\frac{\partial \dot V}{\partial x^{reg}}\frac{\partial x^{reg}_\infty}{\partial V}\right|_{TC} = -\mathlarger{\mathlarger{\sum}}_{x^{r-}}\ \overbrace{\left.\frac{\partial \dot V}{\partial x^{s-}}\frac{\partial x^{s-}_\infty}{\partial V}\right|_{TC}}^{\colorbox{white}{$\mathlarger{<0}$}}} -  \mathlarger{\mathlarger{\sum}}_{x^{s+}\neq x^{reg}}\ \overbrace{\left.\frac{\partial \dot V}{\partial x^{s+}}\frac{\partial x^{s+}_\infty}{\partial V}\right|_{TC}}^{\colorbox{white}{$\mathlarger{>0}$}} $\hfill(b.eq.)\\
\hspace{5mm}(ii-c) If $I_{reg}$ has an adaptation variable $x^a$, pick it as the bifurcation parameter $\lambda$ regulating\\
\hspace{5mm}\hspace{8.5mm} the left hand side of (b.eq.), that is $\lambda=x^a$.\\
\hspace{5mm}\hspace{7.5mm} If $I_{reg}$ has no adaptation variable, pick $\lambda=\bar g_{reg}$.\\
\hline
\noindent(iii) {\bf {Singularity} condition and fixed point equation}\\
\hspace{5mm}(iii-a) Solve (b.eq.) together with the {singularity condition} (\ref{EQ: generic ansatz}), in $V$ and $\lambda$.\\
\hspace{5mm}\hspace{7.5mm} For numerical implementation, recall that the left hand side of (\ref{EQ: generic ansatz}) is proportional to\\
\hspace{0.25\textwidth}{$\displaystyle{\frac{\partial\dot V}{\partial V}+\mathlarger{\mathlarger{\sum}}_{x^f} \frac{\partial\dot V}{\partial x^f} \frac{\partial x^f_\infty}{\partial V}}$}\\
\hspace{5mm}(iii-b) Plug the computed values $V^c$ and $\lambda^c$ into the fixed point equation $\dot V|_{TC}=0$ to\\
\hspace{5mm}\hspace{7.5mm} compute the value of the applied current at the transcritical bifurcation ($I^c$).\\
\hline
\noindent(iv) {\bf Tracking of excitability switches}\\
\hspace{5mm} Change the applied current according to the equation\\
\hspace{0.25\textwidth}$\displaystyle{I_{app}=I^c-\left.\frac{\partial\dot V}{\partial\lambda}\right|_{TC}(\lambda-\lambda^c),}$\\
\hspace{5mm}and compute the model bifurcation diagram with $V$ as the variable and $\lambda$ as the bifurcation parameter.\\
\hline
\end{tabular}
\caption{}\label{TAB: TC algo}
\end{table}

\clearpage

\begin{center}
{\huge Supplementary material}
\end{center}

\newbibliography{supp}
 
\renewcommand{\thesection}{S.\arabic{section}}
\renewcommand{\thesubsection}{\thesection.\arabic{subsection}}
 
%
\makeatletter 
\def\tagform@#1{\maketag@@@{(S\ignorespaces#1\unskip\@@italiccorr)}}
\makeatother
 
\makeatletter
\makeatletter \renewcommand{\fnum@figure}
{\figurename~S\thefigure}
\makeatother




\setcounter{section}{0}
\setcounter{figure}{0}

\renewcommand{\figurename}{Figure}

\section*{Construction of the transcritical bifurcation in generic conductance-based models}

Suppose that, given a bifurcation parameter $\mu\in\{\bar g_i, E_i\}_{i\in\mathcal I}$, there exists a a solution $(V,\mu)=(V_{TC},\mu_{TC})$ satisfying the two degeneracy conditions (7) and (8). Then we claim that, posing $$I_{TC}=\sum_{i}\bar g_i m_{i\infty}^{a_i}(V_{TC}) h_{i\infty}^{b_i}(V_{TC}) (V_{TC}-E_i),$$ which solves the fixed point equation, the model is at a transcritical bifurcation. The existence of the couple $(V_{TC},\mu_{TC})$ in the physiological range is shown for specific examples in the main document.

The center space $E_c$ associated to the degeneracy conditions (7) and (8) is spanned by the vector
\begin{equation*}
v_c=(1,{\mathbf k}),\quad {\mathbf k}:=[k_x]_{x},\quad k_x:=\left.\frac{\partial x_\infty}{\partial V}\right|_{V=V_{TC}},
\end{equation*}
where $x$ runs all fast and recover gating variables in the model, that is ${\mathbf k}$ is the vector whose elements are the slopes of the (in)activation functions of all the fast and recovery gating variables calculated at $V=V_{TC}$.
The associated center manifold $\mathcal M_c$ is exponentially attractive. Indeed, it can easily been shown that, when (7) and (8) are satisfied, the remaining nonzero eigenvalues of the Jacobian of (10) are all negative.
To the first order, the dynamics on $\mathcal M_c$ is given by
\begin{IEEEeqnarray*}{rCl}
C_m\dot V_c&=&-\sum_{i\in\MI}\bar g_i \big(m_{i,\infty}(V_{TC})+k_{m_i}(V_c-V_{TC})\big)^{a_i} \big(h_{i,\infty}(V_{TC})+k_{h_i}(V_c-V_{TC})\big)^{b_i} (V_c-E_i)+I_{app}
\end{IEEEeqnarray*}
Consider the affine reparametrization
\begin{equation}\label{EQ: TC affine reparametrization}
\tilde\mu=\mu,\quad\tilde I_{app}=I_{app}+\left.\frac{\partial\dot V_c}{\partial\mu}\right|_{\substack{\mu=\mu_{TC}\\ V=V_{TC}}}(\mu-\mu_{TC}).
\end{equation}
It is easy to show that, in the affine reparametrization (S\ref{EQ: TC affine reparametrization}), the center manifold dynamics satisfy
\begin{equation*}
\dot V_c=\left.\frac{\partial\dot V_c}{\partial V}\right|_{\substack{\mu=\mu_{TC}\\ V=V_{TC}}}=\left.\frac{\partial\dot V_c}{\partial\tilde\mu}\right|_{\substack{\mu=\mu_{TC}\\ V=V_{TC}}}=0
\end{equation*}
corresponding to the defining condition of a transcritical bifurcation (see \cite[Page 367]{supp}{SEYDEL94}).

\bibliographystyle{supp}{plos2009}
\bibliography{supp}{../../../../refs}{Supplementary references}

%

\end{document}